\documentclass[
 aps, pra,
 amsmath,amssymb,
 10pt,
 final,
tightenlines,
 twoside,
 twocolumn,
 nobibnotes,
nofootinbib,
 superscriptaddress,
showkeys,
 ]
{revtex4-2}

\usepackage[utf8x]{inputenc}
\usepackage[english]{babel}
\usepackage{graphicx}
\usepackage{dcolumn}
\usepackage{bm}
\usepackage{natbib}
\usepackage{float}
\usepackage{hyperref}
\usepackage{slashed}
\usepackage{multirow}
\usepackage{lineno}
\usepackage{booktabs}
\usepackage{silence}
\usepackage{subfigure}
\usepackage{xcolor}
\usepackage{mathtools}

\begin{document}

\selectlanguage{english}



\title{Flavor Enhanced Chromomagnetic Dipole Moment in the Bestest Little Higgs Framework}

\author{\firstname{T.}~\surname{Cisneros-Pérez} }
 \email{tzihue@gmail.com}
  \affiliation{Unidad Acad\'emica de Física, Universidad Autónoma de Zacatecas, Solidaridad, esq. la Bufa S/N, Zacatecas 98060, Mexico}

\author{\firstname{A.}~\surname{Ramirez-Morales}}
 \email{andres.ramirez@tec.mx}
\affiliation{Tecnologico de Monterrey, Escuela de Ingenieria y Ciencias, General Ramon Corona 2514,
Zapopan 45138, Mexico}

\author{\firstname{R.}~\surname{Gamboa-Goni}}
 \email{r.gamboa@tec.mx}
\affiliation{Tecnologico de Monterrey, Escuela de Ingenieria y Ciencias, General Ramon Corona 2514,
Zapopan 45138, Mexico}

\author{\firstname{C.}~\surname{Ortiz}}
 \email{ortizgca@fisica.uaz.edu.mx}
 \affiliation{Unidad Acad\'emica de Física, Universidad Autónoma de Zacatecas, Solidaridad, esq. la Bufa S/N, Zacatecas 98060, Mexico}




\begin{abstract}
We investigate the anomalous Chromomagnetic Dipole Moment (CMDM), $\hat{\mu}_t^{\mathrm{BLHM}}$, of the top quark within the Bestest Little Higgs Model (BLHM). Our study incorporates novel interactions arising from the extended CKM matrix in the BLHM and explores a broad region of the experimentally allowed parameter space, yielding CMDM values on the order of
$10^{-3}$. This result represents an improvement over previous CMDM calculations within the BLHM and makes it competitive with other beyond the Standard Model scenarios. Experimental and model parameter uncertainties are considered and propagated through our calculations, using a Monte Carlo method.

\end{abstract}

\maketitle

\section{INTRODUCTION}

The search for new physics in extensions of the Standard Model (SM) leverages common features shared among various theoretical frameworks, such as extended Higgs mechanisms, extra dimensions, and new symmetries that give rise to additional fields. In this context, we take advantage of the exceptional properties of the top quark to investigate its Chromomagnetic Dipole Moment (CMDM) in the Bestest Little Higgs Model (BLHM)~\cite{schmaltz2010bestest}.

The BLHM and the Type-I Two Higgs Doublet Models (2HDM) offer a rich structure under the broader umbrella of Little Higgs Models (LHM). The BLHM provides a natural and concise solution to challenges such as custodial symmetry violation, divergent singlets, and the mass hierarchy between extended bosons and heavy quarks~\cite{schmaltz2010bestest}. A distinctive feature of the BLHM is its modular structure, involving two separate symmetry-breaking scales, $f$ and $F$, with the hierarchy $F > f$.

Despite its theoretical appeal, the BLHM has received relatively limited attention compared to other LHM 
or beyond the Standard Model (BSM) frameworks, primarily due to the experimental difficulties emerging from probing increasingly high mass scales. Nevertheless, several studies have examined the BLHM in the context of dipole moments~\cite{Cruz-Albaro:2022kty,Cruz-Albaro:2022lks,Cruz-Albaro:2023pah} and rare top decays~\cite{Cisneros-Perez2023}, motivating the calculation of the top quark CMDM within the BLHM. 

Theoretical interest in the CMDM of the top quark, $\hat{\mu}_t$, arises from its large mass and its close connection to heavy particles in several BSM scenarios. The CMDM has been widely studied within both the SM and BSM frameworks. SM calculations~\cite{Martinez:2007qf, Aranda:2020tox, Martinez:2001qs, montano1, tututi2023} predicted values of order $10^{-2}$. BSM contributions to the CMDM have proven to be non-negligible: In Ref.\cite{Martinez:2007qf}, values of $10^{-3}$ were obtained for the Type-II 2HDM, $10^{-2}$ for the top-color-assisted technicolor (TC2) framework, and $10^{-3}$ for 5D models. In the context of the Littlest Higgs Model with T-parity (LHT)\cite{Ding:2008nh}, 2HDM scenarios yielded values in the range $10^{-3}$–$10^{-4}$. For the 331 model~\cite{Hernandez-Juarez:2020gxp}, the CMDM lies between $10^{-7}$ and $10^{-6}$. Decoupling effective field theory analyses~\cite{Martinez:1996cy} set an upper bound of $10^{-2}$. Models with a fourth fermion generation~\cite{Hernandez-Juarez:2018uow} predicted values ranging from $10^{-1}$ to $10^{-2}$. More so, a uniparticle model studied in Ref.\cite{Sampayo:2010} gives a CMDM of order $10^{-2}$, and two-loop 2HDM calculations\cite{Bisal2024} resulted in values between $10^{-4}$ and $10^{-3}$. Finally, within the BLHM, the first computation of the top-quark CMDM was reported in Ref.~\cite{Aranda:2021kza}, yielding values in the range $10^{-4}$–$10^{-6}$.

On the experimental side, the CMS Collaboration, using $pp$ collisions at $\sqrt{s} = 13$~TeV with an integrated luminosity of $35.9$~fb$^{-1}$, has recently reported two independent measurements of the top quark CMDM at the LHC~\cite{Sirunyan2020,Sirunyan2019}. The measurement in Ref.~\cite{Sirunyan2019}, focusing on leptonic final states, yielded a precise value consistent with the SM prediction, while the analysis in Ref.~\cite{Sirunyan2020}, based upon lepton+jets final states, also showed SM consistency, but may offer richer sensitivity to possible deviations, thereby providing a valuable probe for BSM physics:
\begin{equation}\label{cmdmexp}
    \hat{\mu}^{Exp}_t=-0.024^{+0.013}_{-0.009}(\mathrm{stat})^{+0.016}_{-0.011}(\mathrm{syst}).
\end{equation}

In this context, our work updates the magnitude of the top quark's CMDM within the BLHM, via incorporating contributions from new interactions as introduced by Ref.~\cite{Cisneros-Perez2023}, which include flavor-changing processes mediated by two extended matrices, $V_{Hu}$ and $V_{Hd}$, analogous to the Cabibbo–Kobayashi–Maskawa (CKM) matrix~\cite{kobayashi1973cp}. {\color{black}
Furthermore, we evaluate the CMDM over a broad range of BLHM parameters. We first consider the mixing angle $\beta$ in the interval $1.10 \leq \beta \leq 1.40$, consistent with theory and experimental constraints~\cite{schmaltz2010bestest, ATLAS2021tanBeta, ATLAS2024tanBeta, ATLAS2019tanBeta}. Since the Yukawa couplings can significantly affect the CMDM, their entire allowed range is explored. The BLHM also introduces extended symmetry couplings, $g_A$ and $g_B$, hence we study their impact on the CMDM. In the same line, the masses of the heavy quarks in the BLHM depend on the scale $f$, while the masses of the heavy gauge bosons depend on both $f$ and $F$, with the restriction $f+F < 10 $ TeV, due to the cutting scale $\Lambda$, and so, accordingly, the CMDM was explored for several allowed values of $f$. Across these parameter scans, the CMDM is found to lie in the order of $10^{-3}$. This represents an improvement w.r.t. previous CMDM BLHM results, a step forward in the probe of this extension of the SM, since the explored experimental range associated with Eq. \ref{cmdmexp} allows for values of this order.

This paper is organized as follows: Section~\ref{review} provides a brief overview of the BLHM, vital to establish the theoretical framework of our analysis. Section~\ref{chromo} discusses the sector of the BLHM that includes the $gt\bar{t}$ vertex relevant for the $\hat{\mu}^{\text{BLHM}}_{t}$ calculation. In Section~\ref{pspace}, we analyze the allowed phase space involved in the CMDM computation, while Section~\ref{pheno} presents the various parameter scenarios explored, along with our methodology for error propagation. Results are summarized in Section~\ref{results}, including detailed plots illustrating the confidence intervals, which may serve as a guiding posts for future research. To close, conclusions are contained in Section~\ref{conclusions}. Appendix~\ref{feynrules} includes the Feynman rules necessary for computing the CMDM.

\section{Summary of the BLHM}
\label{review}

To naturally include custodial $SU(2)$ symmetry, the BLHM is based on the global symmetry group \mbox{$SO(6)_A \times SO(6)_B$}, which is spontaneously broken to the diagonal subgroup $SO(6)_V$ at the scale $f$. This symmetry breaking occurs when the non-linear sigma field $\Sigma$, also present in other composite Higgs and Little Higgs theories, acquires a vacuum expectation value (VEV), \mbox{$\langle \Sigma \rangle = 1$~\cite{schmaltz2010bestest}}. As a consequence, 15 pseudo-Nambu–Goldstone bosons (pNGBs) emerge. These consist of a zero hypercharge electroweak triplet $\phi_a$ and a triplet $\eta_a$, with $a = 1, 2, 3$, where $(\eta_1, \eta_2)$ is a complex singlet carrying hypercharge, and $\eta_3$ is a real singlet. Finally, there is a real electroweak singlet scalar field $\sigma$ and two quadruplets \mbox{$h_i^T = (h_{i1}, h_{i2}, h_{i3}, h_{i4})$,} with $i = 1, 2$, that transform under the $SO(4)$ subgroup~\cite{Cisneros-Perez2023}. The $h_i$ components are the main contributors to the scalar potential, and their mixing after symmetry breaking gives rise to the scalar fields of the BLHM. With these 15 pNGBs, the sigma field is expressed as
\begin{equation}\label{Sigma}
\Sigma = e^{i\Pi/f} e^{2i\Pi_h/f} e^{i\Pi/f},
\end{equation}
where $\Pi$ is a $6 \times 6$ matrix that contains the fields $\phi_a$, $\eta_a$, and $\sigma$, while $\Pi_h$ comprises the quadruplets $h_i.$


%
%

\subsection{Scalar sector in the BLHM}
In the BLHM, two operators are required to generate the quartic coupling of the Higgs field through collective symmetry breaking. These operators, \mbox{$P_5 = \text{diag}(0, 0, 0, 0, 1, 0)$} and $P_6 = \text{diag}(0, 0, 0, 0, 0, 1)$, need to be applied simultaneously to achieve this 
and develop a Higgs field potential. Consequently, the quartic potential can be expressed as~\cite{schmaltz2010bestest}

\begin{equation}\label{potVq}
 V_q=\frac{1}{4}\lambda_{65}f^4(\Sigma_{65})^2+\frac{1}{4}\lambda_{56}f^4(\Sigma_{56})^2,
\end{equation}

\noindent where $\lambda_{65}$ and $\lambda_{56}$ are coefficients that must be nonzero. Expanding Eq.(\ref{Sigma}) in powers of $1/f$ and substituting the result into Eq.(\ref{potVq}), we obtain

\begin{eqnarray}\label{potVqSerie}
 V_q=\frac{\lambda_{65}}{2}\left(f\sigma-\frac{1}{\sqrt{2}}h_1^Th_2+\dots\right)^2\\\nonumber
 +\frac{\lambda_{56}}{2}\left(f\sigma+\frac{1}{\sqrt{2}}h_1^Th_2+\dots\right)^2.
\end{eqnarray}

\noindent This potential induces a mass term for $\sigma$, given by \mbox{$m_{\sigma}^2 = (\lambda_{65} + \lambda_{56}) f^2$.} To prevent the appearance of an additional quartic coupling for the Higgs fields, one must impose the shift $\sigma \to \pm \frac{h_1^T h_2}{\sqrt{2} f}$ in Eq.(\ref{potVqSerie}). Then, the combined effect of the two terms in Eq.(\ref{potVqSerie}), after integrating out $\sigma$, yields a tree-level quartic potential for the Higgs fields ~\cite{schmaltz2010bestest,Schmaltz:2008vd,phdthesis}:

\begin{equation}
 V_q=\frac{\lambda_{56}\lambda_{65}}{\lambda_{56}+\lambda_{65}}\left(h_1^Th_2 \right)^2=\frac{1}{2}\lambda_{0}\left(h_1^Th_2 \right)^2.
\end{equation}

\noindent In this way, a quartic collective potential, proportional to two independent couplings, is obtained~\cite{schmaltz2010bestest}. Notice that the quartic coupling $\lambda_0$ vanishes if either $\lambda_{56}$ or $\lambda_{65}$ is set to zero, illustrating the mechanism of collective symmetry breaking.

\noindent In the scalar sector, the absence of gauge interactions implies that not all scalar fields acquire a mass. To address this, the following potential is introduced:
\begin{eqnarray}\label{potVs}\nonumber
 V_s&=&-\frac{f^2}{4}m_4^2Tr\left(\Delta^{\dagger}M_{26}\Sigma M^{\dagger}_{26}+\Delta M_{26}\Sigma^{\dagger}M^{\dagger}_{26}\right)\\
 &&-\frac{f^2}{4}\left(m_5^2\Sigma_{55}+m_6^2\Sigma_{66}\right),
\end{eqnarray}
\noindent where $m_4$, $m_5$, and $m_6$ are mass parameters, and $\Sigma_{55}, \Sigma_{66}$ denote components of the matrix $\Sigma$ defined in Eq.~(\ref{Sigma}). The matrix $M_{26}$ serves to contract the $SU(2)$ indices of $\Delta$ with the $SO(6)$ indices of $\Sigma$. The $\Delta$ operator arises from the global symmetry $SU(2)_C \times SU(2)_D$, which is spontaneously broken to its diagonal subgroup $SU(2)$ at the scale $F > f$, when $\Delta$ acquires a vacuum expectation value (VEV), $\langle \Delta \rangle = 1$. It can be parameterized as
\begin{equation}
 \Delta=e^{2i\Pi_d/F},\hspace{0.3cm}\Pi_d=\chi_a\frac{\tau_a}{2}\hspace{0.3cm}(a=1,2,3).
\end{equation}
 The matrix $\Pi_d$ includes the scalars from the $\chi_a$ triplet, which mix with the triplet $\phi_a$, while $\tau_a$ denotes the Pauli matrices. The $\Delta$ operator is related to $\Sigma$ in such a manner that the diagonal subgroup of \mbox{$SU(2)_A \times SU(2)_B \subset SO(6)_A \times SO(6)_B$,} is identified as the SM $SU(2)_L$ group.

Expanding the $\Delta$ operator in powers of $1/F$ and substituting it into Eq. (\ref{potVs}), we obtain
\begin{equation}
 V_s=\frac{1}{2}\left(m^2_{\phi_{a}}\phi^2_a+m^2_{\eta_{a}}\eta^2_a+m^2_1h^T_1h_1+m^2_2h^T_2h_2\right),
\end{equation}
where
\begin{eqnarray}
 m^2_{\phi}&=&m^2_{\eta}=m^2_4,\\\nonumber
 m^2_1&=&\frac{1}{2}(m^2_4+m^2_5),\\\nonumber
 m^2_2&=&\frac{1}{2}(m^2_4+m^2_6).
\end{eqnarray}
To trigger electroweak symmetry breaking (EWSB), the following potential term is introduced~\cite{schmaltz2010bestest}:
\begin{equation}
V_{B_{\mu}} = m_{56}^2 f^2 \Sigma_{56} + m_{65}^2 f^2 \Sigma_{65},
\end{equation}
where the mass parameters $m_{56}$ and $m_{65}$ correspond to the matrix elements $\Sigma_{56}$ and $\Sigma_{65}$, respectively, and
\begin{equation}\label{potB}
B_{\mu} = 2 \frac{\lambda_{56} m_{65}^2 + \lambda_{65} m_{56}^2}{\lambda_{56} + \lambda_{65}},
\end{equation}
is a Higgs-like mass term, related to $h_1^Th_2$. With this contribution, the full scalar potential is given by
\begin{equation}\label{pEscalar}
V = V_q + V_s + V_{B_{\mu}}.
\end{equation}

\noindent To obtain a potential for the Higgs doublets, $V_H$, we minimize Eq.~(\ref{pEscalar}) with respect to $\sigma$ and substitute the resulting expression into Eq.~(\ref{pEscalar}). This yields: 
\begin{eqnarray}\label{potVH}\nonumber
 V_H&=&\frac{1}{2}\Big[m_1^2h_1^Th_1+m_2^2h_2^Th_2-2B_{\mu}h_1^Th_2\\
 &+&\lambda_0(h_1^Th_2)^2\Big],
\end{eqnarray}

\noindent with EWSB requiring that $B_{\mu} > m_1m_2$, and the potential reaching a minimum when $m_1m_2 > 0$. Note that the term $B_{\mu}$ vanishes if $\lambda_{56}$, $\lambda_{65}$, or both, are zero, in accordance with Eq. (\ref{potB}). After EWSB, the Higgs doublets acquire VEVs given by
\begin{equation}\label{aches}
 \langle h_1\rangle=v_1,\hspace{0.3cm}\langle h_2\rangle=v_2.
\end{equation}

\noindent Next, minimization of the potential in Eq.(\ref{potVH}) with respect to the vacuum expectation values in Eq.(\ref{aches}), lead to the following relations:
\begin{eqnarray}
 v_1^2=\frac{1}{\lambda_0}\frac{m_2}{m_1}(B_{\mu}-m_1m_2),\\
 v_2^2=\frac{1}{\lambda_0}\frac{m_1}{m_2}(B_{\mu}-m_1m_2).
\end{eqnarray}
Moreover, the angle $\beta$ between $v_1$ and $v_2$ is defined~\cite{schmaltz2010bestest} by
\begin{equation}
\label{eq:tanbeta}
\tan\beta = \frac{\langle h_{11} \rangle}{\langle h_{21} \rangle} = \frac{v_1}{v_2} = \frac{m_2}{m_1},
\end{equation}
so that the total electroweak scale is given by
\begin{eqnarray}
v^2&=&v_1^2+v_2^2\\\nonumber
&=&\frac{1}{\lambda_0}\left(\frac{m_1^2+m_2^2}{m_1m_2}\right)(B_{\mu}-m_1m_2)\\\nonumber
&\simeq&(246\;\mathrm{GeV})^2.
\end{eqnarray}
\noindent After EWSB, the scalar  sector in the BLHM~\cite{schmaltz2010bestest,phdthesis} gives rise to massive physical states: $h^0$ (identified with the SM Higgs in the BLHM), $A^0$, $H^{\pm}$, and $H^0$, with the following mass spectrum:
\begin{eqnarray}
\label{masa-esc1}
&&m^2_{G^0}=m^2_{G^{\pm}}=0,\\
\label{masa-esc2}
&&m^2_{A^0}=m^2_{H^{\pm}}=m^2_1+m^2_2,\label{masa-A0}\\
\label{masa-esc3}
&&m^2_{H^0,h^0}=\frac{B_{\mu}}{\sin2\beta}\label{masa-H0}\\\nonumber
\pm&&\sqrt{\frac{B^2_{\mu}}{\sin^22\beta}-2\lambda_0B_{\mu} v^2\sin2\beta+\lambda_0^2v^4\sin^22\beta}.
\end{eqnarray}
\noindent $G^0$ and $G^{\pm}$ refer to the absorbed Goldstone bosons that give mass to the SM $Z$ and $W^{\pm}$ gauge bosons.


\subsection{Gauge sector in the BLHM}
The gauge terms in the BLHM are provided by the Lagrangian~\citep{schmaltz2010bestest,phdthesis},
\begin{equation}\label{lag-norma}
\mathcal{L}=\frac{f^2}{8}Tr\left(D_{\mu}\Sigma^{\dag}D^{\mu}\Sigma\right)+\frac{F^2}{4}Tr\left(D_{\mu}\Delta^{\dag}D^{\mu}\Delta\right),
\end{equation}
where \( D_{\mu} \Sigma \) and \( D_{\mu} \Delta \) are covariant derivatives of the form
\begin{eqnarray}
 D_{\mu}\Sigma&=&i\sum_a\left(g_AA_{1\mu}^aT_L^a\Sigma-g_BA_{2\mu}^a\Sigma T^a_L\right)\\\nonumber
 &+&ig'B_3\left(T^3_R\Sigma-\Sigma T^3_R\right),\\
 D_{\mu}\Delta&=&\frac{i}{2}\sum_a\left(g_AA^a_{1\mu}\tau_a\Delta-g_BA^a_{2\mu}\Delta\tau_a\right).
 \end{eqnarray}
Here, \( A^a_{1\mu} \) and \( A^a_{2\mu} \) represent the gauge boson eigenstates. 
{ $T_L^a$ and $T_R^a$ are the generators of the $SU(2)_L$ and $SU(2)_R$ groups \cite{phdthesis}, respectively, which together form the $SO(4)$ group via the product $SU(2)_L \times SU(2)_R$.}
The coupling \( g' \) corresponds to the $U(1)_Y$ gauge group, while \( g \) is associated with $SU(2)_L$. 
These couplings, \( g' \) and \( g \), are related to the couplings \( g_A \) and \( g_B \) of the extended symmetry group \( SU(2)_A \times SU(2)_B \) through the following relations:

\begin{eqnarray}\label{acoplesSU}
g&&=\frac{g_Ag_B}{\sqrt{g_A^2+g_B^2}},\\
s_g&&=\sin\theta_g=\frac{g_A}{\sqrt{g_A^2+g_B^2}},\\
\label{acoplesSU1}
c_g&&=\cos\theta_g=\frac{g_B}{\sqrt{g_A^2+g_B^2}},
\end{eqnarray}
with $\theta_g$ being the mixing angle.

Finally, the masses of the heavy gauge bosons \( W^{\prime \pm} \), \( Z' \), in the BLHM, along with those of the SM gauge bosons, are also generated via the EWSB mechanism~\citep{schmaltz2010bestest, phdthesis},
\begin{eqnarray}
\label{masa-zp}
m_{Z'}^2&=&\frac{1}{4}(g_A^2+g_B^2)(f^2+F^2)-\frac{1}{4}g^2v^2,
\end{eqnarray}
\begin{eqnarray}
\label{masa-wp}
m_{W'^{\pm}}^2&=&\frac{1}{4}(g_A^2+g_B^2)(f^2+F^2)-m_{W^{\pm}}^2.
\end{eqnarray}


\subsection{Fermion sector  in the BLHM}

The structure of the fermion sector is given by the Lagrangian~\cite{schmaltz2010bestest}
\begin{eqnarray}
\label{lag-yuk}
\mathcal{L}_t&=&y_1fQ^TS\,\Sigma\, SU^c+y_2fQ_a^{\prime T}\Sigma\,U^c\\\nonumber
&+&y_3fQ^T\Sigma\, U_5^{\prime c}+y_bfq_3^T(-2iT_ R^3\Sigma)U_b^c+\textrm{h.c.},
\end{eqnarray}

\noindent where $Q,Q^{\prime}_a,q_3$ and $U^c,U^{\prime c}_5,U^c_b$ are multiplets of $SO(6)_A$ and $SO(6)_B$, respectively ~\cite{phdthesis}, \mbox{$S=diag(1,1,1,1,-1,-1)$} is a symmetry operator, $y_1,y_2,y_3$ are Yukawa couplings, and the terms $q_3$ and $U_b^c$, contain information about the bottom quark.

Most importantly, top quark loops contribute the largest divergent quantum corrections to the Higgs mass in the SM, hence the new heavy quarks in the BLHM framework play a key role in addressing the hierarchy problem. Those heavy quarks are: $T$, $T^5$, $T^6$, $T^{2/3}$, $T^{5/3}$, and $B$~\cite{schmaltz2010bestest}.

In the quark sector Lagrangian, the Yukawa couplings must satisfy $0<y_i<1$. The expresion for the quark top mass contains its Yukawa coupling $y_t$~\cite{phdthesis}, such that:
\begin{equation}
m^2_t=y_t^2v_1^2\label{masa-top}.
\end{equation}

\noindent The coupling $y_t$ is given by:
\begin{equation}\label{acople-yt}
 y_t^2=\frac{9y_1^2y_2^2y_3^2}{(y_1^2+y_2^2)(y_1^2+y_3^2)},
\end{equation}
that is part of the measure of fine-tuning, $\Psi$, in the BLHM \cite{phdthesis},
\begin{equation}
\label{ajuste-fino}
\Psi=\frac{27f^2}{8\pi^2v^2\lambda_0\cos^2\beta}\frac{|y_1|^2|y_2|^2|y_3|^2}{|y_2|^2-|y_3|^2}\log\frac{|y_1|^2+|y_2|^2}{|y_1|^2+|y_3|^2}.
\end{equation}
$\Psi$ serves as measure to quantify how sensitive the electroweak scale is to radiative corrections in different sectors~\cite{schmaltz2010bestest}.

\subsection{Flavor mixing in the BLHM}

The authors of Ref.~\cite{Aranda:2021kza} considered only the contributions arising from interactions between the heavy top quarks of the BLHM and the SM top quark, mediated by the bosons present in both theories. 
By introducing new interaction terms between the heavy top quarks and the SM light quarks, interactions not present in the original BLHM, we can extend the phenomenology of the model and enhance the predicted magnitude of the CMDM. 
To achieve this, we adopt the flavor structure introduced in Ref.~\cite{Cisneros-Perez2023} for the BLHM. The latter preserves the model's fundamental symmetries and leaves the coupling vertices between heavy top-like quarks and SM quarks unchanged. The modification to the BLHM,  in Ref.~\cite{Cisneros-Perez2023}, consists of adding terms to the Lagrangians that describe interactions among the fields $(W^{\prime\pm},H^{\pm},\phi^{\pm},\eta^{\pm})$, the heavy quark $B$, and the light SM quarks $(u, c, d, s)$, while avoiding tree-level FCNCs. Thus, for scalar interactions, the terms
\begin{equation}
    \label{eq:yb}
y_Bfq_1(-2iT_R^2\Sigma)d_B^c,\hspace{0.2cm}y_Bfq_2(-2iT_R^2\Sigma)d_B^c,
\end{equation}
are added to the Lagrangian in Eq.(\ref{lag-yuk}). In Eq.~(\ref{eq:yb}), the relation \( y_B^2 = (y_1^2 + y_2^2)/2 \) defines the Yukawa coupling for the heavy \( B \) quark, where \( y_B < 1 \). Here, \( q_1 \) and \( q_2 \) represent multiplets of light SM quarks, while \( d_B^c \) is a new multiplet containing the \( B \) quark. These fields are expressed as follows:
{
\begin{eqnarray}\label{q-nuevo}
    q_1^T&=&\frac{1}{\sqrt{2}}(-u,iu,d,id,0,0),\\\nonumber
    q_2^T&=&\frac{1}{\sqrt{2}}(-c,ic,s,is,0,0),\\\nonumber
    d_B^{c\,T}&=&(0,0,0,0,B,0).
\end{eqnarray}

For the vector interactions among the fields $W^{\pm},W'^{\pm}$, the heavy quark $B$, and the light SM quarks, the terms 
\begin{equation}
    \sum_{i=1}^2i\bar{\sigma}_{\mu}Q^{\dagger}_3D^{\mu}q_i,\hspace{0.5cm}\sum_{i=1}^4i\bar{\sigma}_{\mu}q_i^{\prime\dagger}D^{\mu}U^c,
\end{equation}
are added to the fermion-gauge Lagrangian in the BLHM~\cite{schmaltz2010bestest,phdthesis}, where $\bar\sigma_{\mu}$ are the Pauli matrices. The multiplet $Q_3$ contains the quark $B$, and the multiplets $q'_i$ represent the light quarks $(u, c, d, s)$. The $Q_3$ and $q'_i$ are given by
\begin{eqnarray}
    Q_3^T&=&\frac{1}{\sqrt{2}}(0,0,B,iB,0,0),\\\nonumber
    q_i^{\prime T}&=&(0,0,0,0,q_i^c,0).
\end{eqnarray}
Using these definitions, we write the extended Lagrangian for fermion-gauge interactions as~\cite{Cisneros-Perez2023}:
    \begin{eqnarray}\label{lag-Q}
    \mathcal{L}&=&i\sum_{i=1}^2\bar{\Psi}_{q_i}\gamma_{\mu}P_LD^{\mu}\Psi_{q_i}+i\bar{\Psi}_{Q}\gamma_{\mu}P_LD^{\mu}\Psi_{Q}\\\nonumber
    &+&\bar{\Psi}_{Q'}\gamma_{\mu}P_LD^{\mu}\Psi_{Q'}+i\bar{\Psi}_{U^c}\gamma_{\mu}P_RD^{\mu}\Psi_{U^c}\\\nonumber
    &+&i\sum_{i=1}^2\bar{\Psi}_{Q_3}V_H^{\dagger}\gamma_{\mu}P_LD^{\mu}\Psi_{q_i}+i\sum_{i=1}^4\bar{\Psi}_{q^{\prime}_i}\gamma_{\mu}P_RD^{\mu}V_H\Psi_{U^c},
    \end{eqnarray}
     Here, $\Psi_X$ denotes the Dirac spinor representation of the quark field $X$ and the left- and right-handed projection operators defined as $P_L = \frac{1}{2}(I - \gamma^5)$ and $P_R = \frac{1}{2}(I + \gamma^5)$, respectively. Using these definitions, the flavor states can be expressed as rotated states via the extended CKM matrix:
    \begin{equation}
 \Psi_{q_i}=
 \begin{pmatrix}
  V_{u}\\
  V_{d}
 \end{pmatrix}\Psi^{\prime}_{q_i},\hspace{0.5cm}
 \Psi_{q^{\prime}_i}=
 \begin{pmatrix}
  V_{u}\\
  V_{d}
 \end{pmatrix}\Psi^{\prime}_{q^{\prime}_i},
\end{equation}
where the extended CKM matrix is such that $V_H^{\dagger}V_u=V_{Hu}$, $V_H^{\dagger}V_d=V_{Hd}$.}
    All covariant derivatives contain gauge bosons $W^{\pm}$ and $W^{\prime\pm}$ \cite{phdthesis,Martin:2012kqb}, and take the form:
    \begin{equation}
        D^{\mu}\Psi_X=\left(\partial^{\mu}+i\sum_{a=1}^2g_AA_1^{a\mu}T^a_L\right)\Psi_X,
    \end{equation}
    where, $A_1^{a\mu}$~\cite{phdthesis} explicitly contains the fields $W^{\pm},W^{\prime\pm}$ so that we can write for $\Psi_Q$ from Eq. \ref{lag-Q} first line:
    \begin{eqnarray}\nonumber
        &&\frac{ig_A}{\sqrt{2}}\bar{\Psi}_Q\gamma_{\mu}P_L\Big([\rho_{11}(W^{+\mu}+W^{-\mu})+\delta_{11}(W^{\prime+\mu}+W^{\prime-\mu})]\\
        &+&[\rho_{12}(W^{+\mu}-W^{-\mu})+\delta_{12}(W^{\prime+\mu}-W^{\prime-\mu})]\Big)\Psi_Q,
    \end{eqnarray}
    where $(\rho_{11},\rho_{12})$ and $(\delta_{11},\delta_{12})$ encompass model constants and a dependency on $\mathcal{O}(v^2/(f^2+F^2))$. The multiplet $Q$ contains the quark $B$, hence we  write for $W^{\prime-}$:
    \begin{equation}
        \frac{ig_A}{\sqrt{2}}\bar{\Psi}_BV_{Bt}\gamma_{\mu}P_LW^{\prime-}\Psi_t,
    \end{equation}
    where $V_{Bt}\in V_{Hu}$. The same procedure applies to $V_{Hd}$.
These contributions made to the BLHM in \cite{Cisneros-Perez2023} extend the phenomenology provided by the model in such a way that two CKM-like unitary matrices, $V_{Hu}$ and $V_{Hd}$, can now be associated, satisfying the relation $V_{CKM}=V_{Hu}^{\dagger}V_{Hd}$, where $V_{CKM}$ is the Cabibbo-Kobayashi-Maskawa matrix \cite{kobayashi1973cp}.

\begin{figure*}[ht]
\includegraphics[scale=0.5]{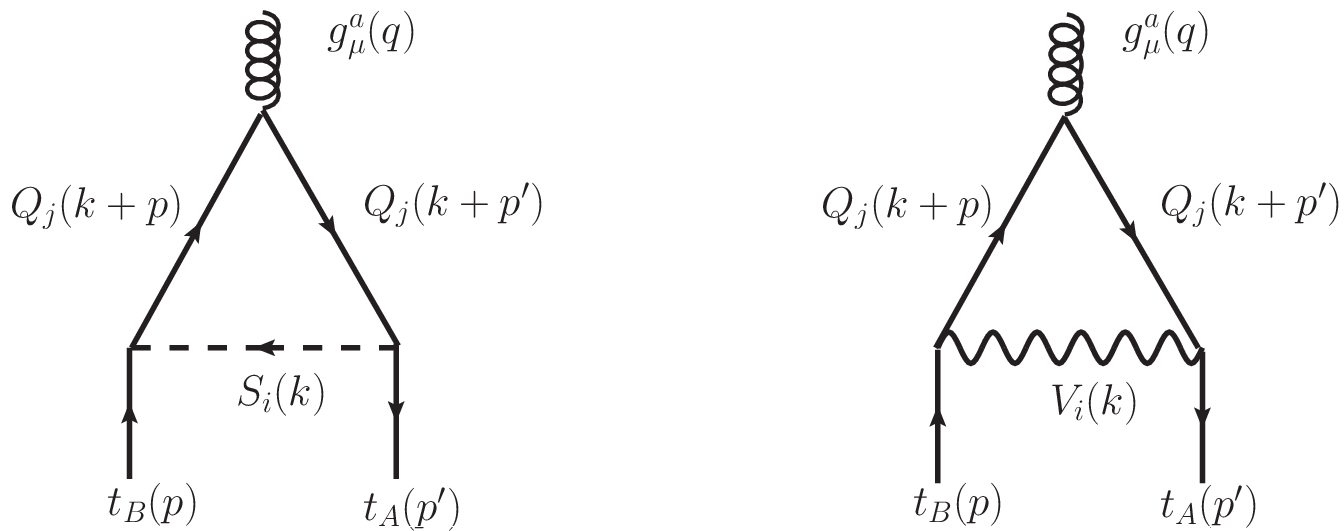}
\caption{The left diagram shows the interactions of the SM top quark and the BLHM heavy quarks \mbox{$Q_j=(T,T^5,T^6,T^{2/3},T^{5/3},B)$,} with the scalar fields $S_i=(A^0, H^0, h^0, H^{\pm}, \phi^0, \eta^0, \sigma, \phi^{\pm}, \eta^{\pm})$. The right diagram shows the interactions of the SM top quark and $Q_j$ with the vector fields $V_i=(Z^0, W^{\pm}, \gamma, Z^{\prime}, W^{\prime\pm}$). Both diagrams also include the SM top and bottom quarks in the loop. $q$ represents the initial gluon momentum, $k$ the loop momentum, and $p$ and $p'$ the external momentum of $t$ and $\bar{t}$, respectively. $A$ and $B$ are color indexes.
}
 \label{dipolo}
\end{figure*}

\section{The chromomagnetic dipole moment of the top quark in the BLHM}
\label{chromo}

The search for new physics can be pursued through the study of anomalous magnetic moments which arise in systems involving fermion pairs, such as electrons and muons, or in more complex composite systems, like with the neutron. Top quark's anomalous magnetic moment is a powerful probe of SM and BSM predictions. Specifically, the top quark high mass and strong Higgs sector coupling, allows for BSM virtual loop correction scenarios enhancements to occur, such that if the considered process actually exists in nature, the former may fall within the scope of experimental sensitivity, due to this, thus offering indirect access to energy scales potentially inaccessible to direct detection. Moreover, the small quark lifetime and lack of hadronization, enables the possibility to experimentally study it's spin properties at high energy particle colliders, also allowing the probing for BSM physics.

It is well-known that the effective Lagrangian is non-renormalizable for the CMDM; however, it can still parameterize sufficiently large contributions to be detectable at the LHC. Formally, the $g\,\bar{t}t$ vertices in \mbox{Fig. \ref{dipolo}} are used to compute one-loop corrections of the top quark's chromomagnetic dipole moment, enabling the introduction of a five-dimensional operator to parameterize the contributions to the CMDM from the BLHM. This is done via the Lagrangian,
\begin{equation}
    \mathcal{L}_{eff}=-\frac{1}{2}\bar{t}\sigma^{\mu\nu}\left(\hat{\mu}_t+i\hat{d}_t\gamma^5\right)tG^a_{\mu\nu}T^a,
\end{equation}

\noindent where $G^a_{\mu\nu}$ is the gluon strength tensor, $T^a$ are the $SU(3)$ generators, $\hat{\mu}_t$ is the CMDM, and $\hat{d}_t$ is the Chromoelectric Dipole Moment (CEDM), such that

\begin{equation}\label{dip-usual}
    \hat{\mu}_t=\frac{m_t}{g_s}\mu_t,\hspace{0.5cm}\hat{d}_t=\frac{m_t}{g_s}d_t.
\end{equation}
The definitions introduced in Eq.~(\ref{dip-usual}) correspond to the standard expressions for the CMDM and the CEDM found in the literature. Here, \(m_t\) denotes the top quark mass, and \(g_s = \sqrt{4\pi \alpha_s}\) is the strong coupling constant. In our analysis, we focus exclusively on computing the chromomagnetic form factor \(\mu_t\), which receives one-loop contributions from the scalar fields \mbox{$S_i=(A^0, H^0, h^0, H^{\pm}, \phi^0, \eta^0, \sigma, \phi^{\pm}, \eta^{\pm})$}, the vector fields \mbox{$V_i=(Z^0, W^{\pm}, \gamma, Z^{\prime}, W^{\prime\pm})$}, and the heavy quarks \mbox{$Q_j=(T,T^5,T^6,T^{2/3},T^{5/3},B)$}.

%
In the context of the CMDM, the valid one-loop diagrams with scalar and vector contributions are shown in Fig. \ref{dipolo}. The amplitudes corresponding to the diagrams are given by:
\begin{eqnarray}\label{ampE}
  &&\mathcal{M}^{\mu}_t(S_i)=\sum_{j}\int\frac{d^4k}{(2\pi)^4}\bar{u}(p^{\prime})(S^{\ast}_i+P^{\ast}_i\gamma^5)\delta_{A\alpha_1}\\\nonumber
  &&\times\left[i\frac{\slashed{k}+\slashed{p}^{\prime}+m_{Q_j}}{(k+p^{\prime})^2-m_{Q_j}^2}\delta_{\alpha_1\alpha_3}\right]\left(-ig_s\gamma^{\mu}T^a_{\alpha_2\alpha_3}\right)\\\nonumber
  &&\times\left[i\frac{\slashed{k}+\slashed{p}+m_{Q_j}}{(k+p)^2-m^2_{Q_j}}\right](S_i+P_i\gamma^5)\delta_{B\alpha_4}V_{Ht}^{\ast}V_{Ht}u(p)\\\nonumber
  &&\times\left(\frac{i}{k^2-m^2_{S_i}}\right),
\end{eqnarray}
\noindent and
\begin{eqnarray}\label{ampV}
  &&\mathcal{M}^{\mu}_t(V_i)=\sum_{j}\int\frac{d^4k}{(2\pi)^4}\bar{u}(p^{\prime})\gamma^{a_1}(V^{\ast}_i+A^{\ast}_i\gamma^5)\delta_{A\alpha_1}\\\nonumber
  &&\times\left[i\frac{\slashed{k}+\slashed{p}^{\prime}+m_{Q_j}}{(k+p^{\prime})^2-m_{Q_j}^2}\delta_{\alpha_1\alpha_3}\right]\left(-ig_s\gamma^{\mu}T^a_{\alpha_2\alpha_3}\right)\\\nonumber
  &&\times\left[i\frac{\slashed{k}+\slashed{p}+m_{Q_j}}{(k+p)^2-m^2_{Q_j}}\right]\gamma^{a_2}(V_i+A_i\gamma^5)\delta_{B\alpha_4}V^{\ast}_{Ht}V_{Ht}u(p)\\\nonumber
  &&\times\left[\frac{i}{k^2-m^2_{V_i}}\left(-g_{\alpha_1\alpha_2}+\frac{k_{\alpha_1}k_{\alpha_2}}{m^2_{V_i}}\right)\right],
\end{eqnarray}

\noindent where $T^a_{\alpha_n\alpha_m}$ are the $SU(3)$ generators, and $A$, $B$, $\alpha_{n,m}$ are the color indexes. $q$ represents the initial gluon momentum, $k$ the loop momentum, and $p$ and $p'$ the external momentum of $t$ and $\bar{t}$, respectively. The $(S_i,P_i,V_i,A_i)$ coefficients carry all contributions from the BLHM, quantified by the vertices $\bar{Q}_jS_it$, $\bar{t}S_i^{\dagger}Q_j$ for scalar and pseudoscalar interactions, and $\bar{Q}_jV_it$, $\bar{t}V_i^{\dagger}Q_j$ for vector and axial interactions, respectively. The matrix elements $V^{\ast}_{Ht}V_{Ht}$ belong to the extended CKM matrix $V_{Hu}$.

Starting from the amplitudes in Eqs.~(\ref{ampE})–(\ref{ampV}), we compute the magnetic form factor $F_2$. 
Then, the BLHM correction to the top quark's CMDM is extracted from $F_2$, yielding the observable $\mu_t$, 
which in turn is substituted into the first relation in Eq.~(\ref{dip-usual}).

For interactions involving charged scalar and vector bosons, we consider the extended CKM matrix for the BLHM, given by $V_{\text{CKM}} = V_{Hu}^{\dagger} V_{Hd}$, as introduced in Ref.~\cite{Cisneros-Perez2023}. 
Here, the unitary matrix $V_{Hu}^{\dagger}$ describes transitions from heavy quarks to light up-type quarks, while $V_{Hd}$ describes transitions from heavy quarks to light down-type quarks. This extended CKM matrix can be generalized as the product of three rotation matrices, following the formalism in Refs.~\cite{blanke2007another,blanke2007rare},

\begin{eqnarray}\label{VHd}
V_{Hd}\;&&=\begin{pmatrix}
1&0&0\\
0&c_{23}^d&s_{23}e^{-i\delta_{23}^d}\\
0&-s_{23}^de^{i\delta_{23}^d} &c_{23}^d
\end{pmatrix}\\\nonumber
&&\times \begin{pmatrix}
c_{13}^d&0&s_{13}^de^{-i\delta_{13}^d}\\
0&1&0\\
-s_{13}e^{i\delta_{13}^d}&0&c_{13}^d
\end{pmatrix}\\\nonumber
&&\times \begin{pmatrix}
c_{12}^d&s_{12}^de^{-i\delta_{12}^d}&0\\
-s_{12}^de^{i\delta_{12}^d}&c_{12}^d&0\\
0&0&1
\end{pmatrix},
\end{eqnarray}

\noindent where the $c^d_{ij}$ and $s^d_{ij}$ entries are in terms of the flavor-mixing angles $(\theta_{12},\theta_{23},\theta_{13})$, and the phases $(\delta_{12},\delta_{23},\delta_{13})$.

\section{Parameter space of the BLHM}
\label{pspace}
M. Schmaltz \textit{ et al.} \cite{schmaltz2010bestest}, considered the BLHM for a Higgs mass between $115$ GeV and $250$ GeV. In subsequent BLHM publications \cite{phdthesis,Martin:2012kqb,Godfrey:2012tf,kalyniak2015constraining}, the Yukawa couplings $y_1, y_2, y_3$ were parameterized keeping $m_{h^0}$ around \mbox{125 GeV.} Additionally, the authors of Ref.~\cite{Cisneros-Perez2023} explored an alternative parameter space to optimize and constrain the BLHM in light of updated experimental data.

Similarly, in this work, we consider a parameter space defined in terms of the mixing angle \(\beta\), which relates the vacuum expectation values \(v_1\) and \(v_2\) as defined in Eq.~(\ref{eq:tanbeta}). Experimental constraints restrict the range of \(\tan\beta\) to \([2, 60]\)~\cite{ATLAS2021tanBeta, ATLAS2024tanBeta, ATLAS2019tanBeta}. Furthermore, due to the structure of the Yukawa sector in the BLHM, Ref.~\cite{schmaltz2010bestest} imposes the condition \(\tan\beta > 1\) in order to suppress one-loop radiative corrections from the top quark and heavy top partners to the Higgs mass. Additionally, the fine-tuning parameter $\Psi$ can take values in the range $0 < \Psi < 10$, with $\Psi \sim 5$ corresponding to a moderate tuning of approximately $20\%$. Moreover, we maintain the Yukawa couplings of the BLHM  in the range $0<y_i<1$, ensuring that the relation in Eq.(\ref{acople-yt}) is satisfied under the condition in Eq.(\ref{masa-top}). These considerations restrict the mixing angle to the range \(1.1 \leq \beta \leq 1.4\) radians. 

We set the condition $\lambda_0<4\pi$ for the quartic coupling. To determine the value of $B_{\mu}$ we use a more suitable variation of Eq. (\ref{potB})\cite{kalyniak2015constraining},

\begin{equation}\label{eq:Bmu}
B_{\mu} = \frac{1}{2}\left( \lambda_0 v^2 + m_{A^{0}}^2 \right) \sin{2\beta}.
\end{equation}

\noindent Moreover, the mixing angle $\alpha$ between $h^0$ and $H^0$, is determined through the relation \cite{phdthesis},
\begin{eqnarray}
\label{ang-alfa}
&&\tan\alpha=\frac{1}{B_{\mu}-\lambda_0 v^2\sin2\beta}\Bigg(B_{\mu}\cot2\beta\\\nonumber
&+&\sqrt{\frac{B_{\mu}^2}{\sin^22\beta}-2\lambda_0 B_{\mu}v^2\sin2\beta+\lambda_0^2 v^4\sin^22\beta}\Bigg).
\end{eqnarray}

The masses of the $A^0$, $H^0$ and $H^{\pm}$ scalar bosons are given by Eq.(\ref{masa-A0})-(\ref{masa-H0}). The mass $\sigma$ boson, is given by $m_{\sigma}^2=2\lambda_0K_{\sigma}f^2,$ where $K_{\sigma}$ is a free parameter of the BLHM~\cite{kalyniak2015constraining}. For the scalar boson $\eta^0$, we use $ m_{\eta^0}=m_4$, where $m_4$ is a free parameter of the model \cite{schmaltz2010bestest}, we choose $m_4 = 500$ GeV \cite{Cisneros-Perez2023}  due to the growing magnitudes of the masses for the new particles that have been experimentally sought. In the case of the charged scalar bosons $\phi^{\pm}$, $\eta^{\pm}$, and the neutral scalar $\phi^0$, their masses depend not only on $m_4$, but also on both symmetry breaking scales, $f$ and $F$, as well as on one-loop contributions from the Coleman–Weinberg potential~\cite{schmaltz2010bestest,Martin:2012kqb}. 

The BLHM is constructed to be valid up to energies of 10 TeV, as it incorporates two symmetry-breaking scales, $f$ and $F$, with the condition $f < F$. In accordance with experimental searches for heavy vector bosons, whose masses are currently investigated around 1 TeV and 7 TeV~\cite{RAPPOCCIO2019100027}, we consider the range $1 \leq f \leq 3$ TeV. Since $F$ is arbitrarily larger than $f$ but is restricted by the model to lie within $3 \leq F \leq 10$ TeV, we adopt this range to ensure that the masses of the heavy gauge bosons $Z'$ and $W'$, given by Eqs.~(\ref{masa-zp}) and (\ref{masa-wp}), remain experimentally accessible with current collider capabilities.

The BLHM introduces five heavy up-type quarks and one heavy bottom-type quark. The masses of the heavy quarks depend on the Yukawa couplings $(y_1, y_2, y_3)$ and the scale~$f$.  
For the angle~$\beta$, we observe only a weak dependence, with effects below~5\% for $T^5$. Furthermore, we set $m_{T^{5/3}}=m_{T^6}=m_{T^{2/3}}$ in the range $550$-$1650$ GeV.

\begin{table*}[htbp]
\caption {Constrained parameters and particle masses in the Bestest Little Higgs Model at energy scales of 1 TeV, 2 TeV, and 3 TeV. The ranges $\beta_{min}=1.1 \leq \beta \leq \beta_{max}=1.4$ radians are shown for all quantities.} \label{masas-esc1}
\medskip
\begin{tabular}{c|c|c|c|c|c|c|c}
\hline
\hline
\multirow{2}{*}{Parameter } & \multicolumn{2}{c}{$f(1\,{\scriptstyle\text{TeV}})$}  & \multicolumn{2}{c}{$f(2\,{\scriptstyle\text{TeV}})$} & \multicolumn{2}{c}{$f(3\,{\scriptstyle\text{TeV}})$}&\\
\cline{2-8}
  & \hspace{3mm}$\beta_{min}$\hspace{3mm} & \hspace{3mm}$\beta_{max}$\hspace{3mm} & \hspace{3mm}$\beta_{min}$\hspace{3mm} & \hspace{3mm}$\beta_{max}$\hspace{3mm} & \hspace{3mm}$\beta_{min}$\hspace{3mm}  & \hspace{3mm}$\beta_{max}$\hspace{3mm} & \hspace{3mm}Unit\hspace{3mm}  \\
\hline
$\beta$            & 1.1       & 1.4        & 1.1       & 1.4       & 1.1      & 1.4     & rad\\
$\alpha$           & -0.456    & -0.111     & -0.456    & -0.111    & -0.456   & -0.111  & rad\\
$y_3$              & 0.37      & 0.32       & 0.37      & 0.32      & 0.37     & 0.32    & rad\\
$\Psi$             & 0.102     & 0.57       & 0.41      & 2.26      & 0.92     & 5.09    & -- \\
$B_\mu$            & 317189    & 155336     & 317189    & 155336    & 317189   & 155336  & GeV$^{2}$ \\
$m_{A^0}$          & 155.47    & 408.59     & 155.47    & 408.59    & 155.47   & 408.59  & GeV\\
$m_{H^0}$          & 876.93    & 954.87     & 876.93    & 954.87    & 876.93   & 954.87  & GeV\\
$m_{H^{\pm}}$      & 155.47    & 408.59     & 155.47    & 408.59    & 155.47   & 237.95  & GeV\\
$m_{\sigma}$       & 5010.00   & 5010.00    & 10030     & 10030     & 15040    & 15040   & GeV\\
$m_{\phi^0}$       & 4487.31   & 4487.31    & 4455.7    & 4455.7    & 4423.77  & 4423.77 & GeV\\
$m_{\phi^{\pm}}$   & 4490.08   & 4490.08    & 4464.3    & 4464.3    & 4437.83  & 4437.83 & GeV\\
$m_{\eta^0}$       & 500.00    & 500.00     & 500.00    & 500.00    & 500.00   & 500.00  & GeV\\
$m_{\eta^{\pm}}$   & 506.48    & 506.48     & 525.45    & 525.45    & 555.63   & 555.63  & GeV\\
$m_{Z'}$           & 4938.55   & 4938.55    & 5627.1    & 5627.1    & 6617.35  & 6617.35 & GeV\\
$m_{W'}$           & 4938.55   & 4938.55    & 5627.1    & 5627.1    & 6617.35  & 6617.35 & GeV\\
$m_T$              & 1063.95   & 1062.87    & 2114.13   & 2113.53   & 3167.33  & 3166.97  & GeV\\
$m_{T^{5}}$        & 625.36    & 599.71     & 1306.52   & 1256.88   & 1900.86  & 1978.63  & GeV\\
$m_{T^{6}}$        & 550.00    & 550.00     & 1100.00   & 1100.00   & 1650.00  & 1650.00  & GeV\\
$m_{T^{2/3}}$      & 550.00    & 550.00     & 1100.00   & 1100.00   & 1650.00  & 1650.00  & GeV\\
$m_{T^{5/3}}$      & 550.00    & 550.00     & 1100.00   & 1100.00   & 1650.00  & 1650.00  & GeV\\
$m_{B}$            & 1140.18   & 1140.18    & 2280.35   & 2280.35   & 3420.53  & 3420.53  & GeV\\
\hline
\hline
\end{tabular}
\end{table*}

\begin{figure*}[htbp]
  \subfigure[]{
    \includegraphics[width=0.485\textwidth]{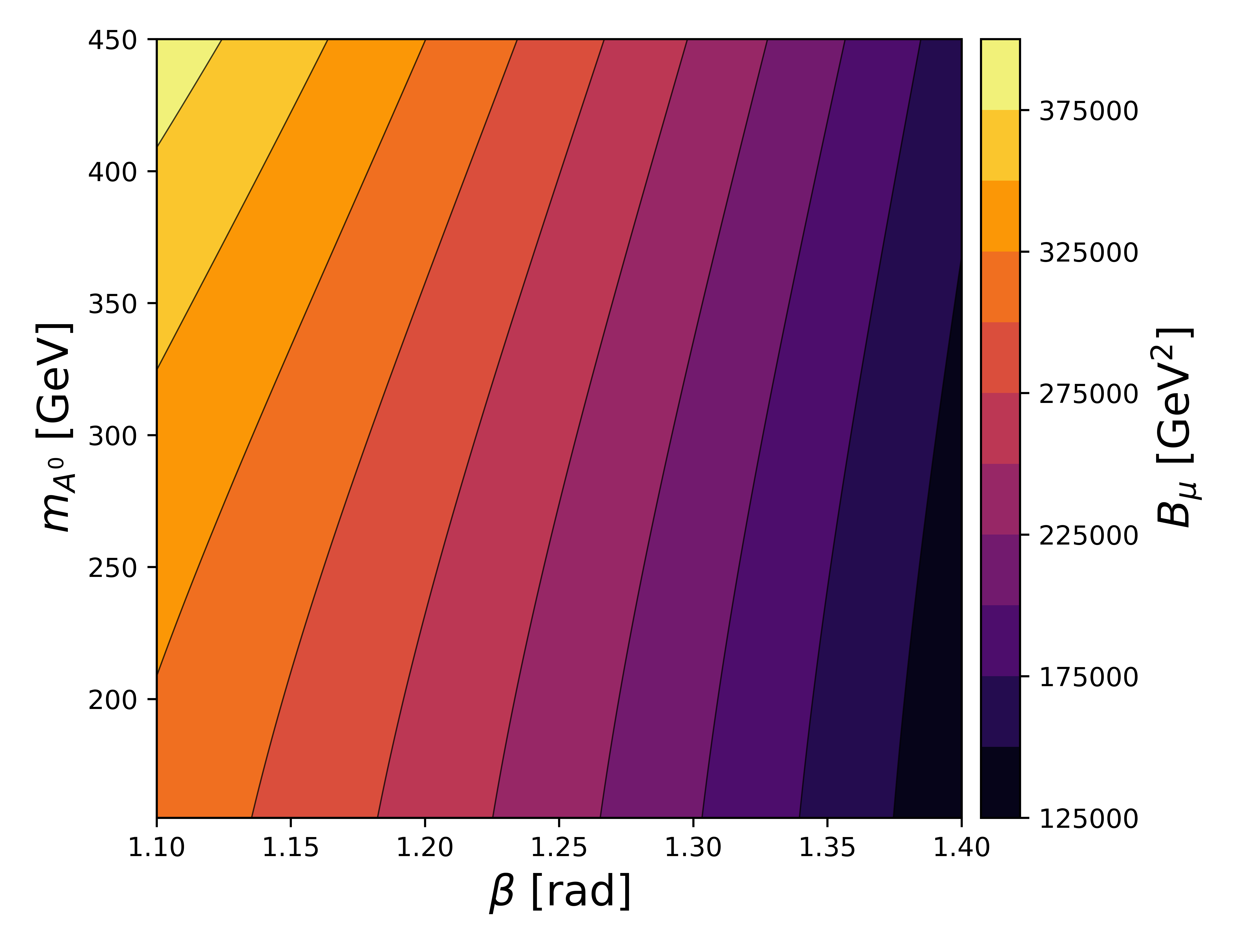}
  }
  \subfigure[]{
    \includegraphics[width=0.485\textwidth]{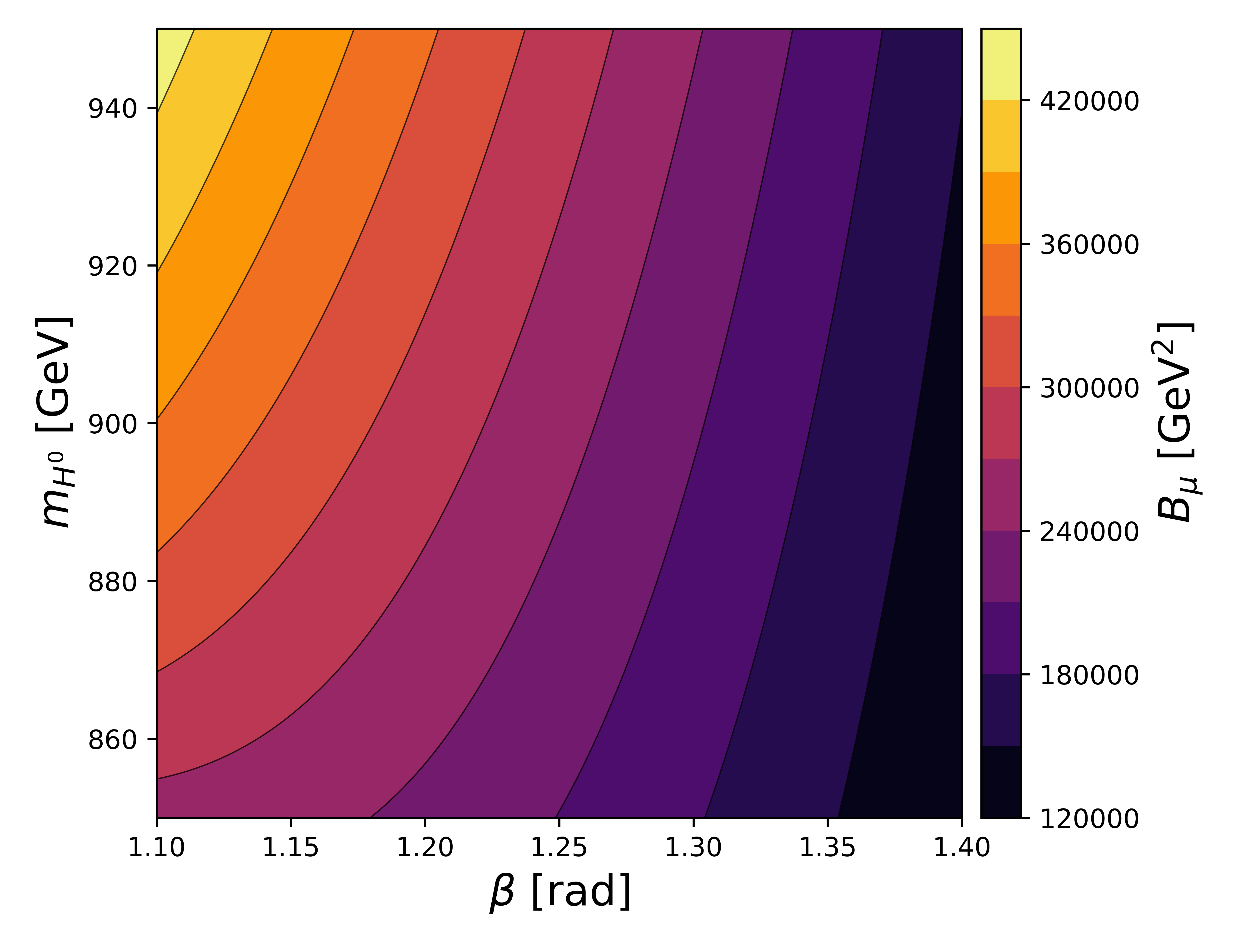}
  }
  \subfigure[]{
    \includegraphics[width=0.485\textwidth]{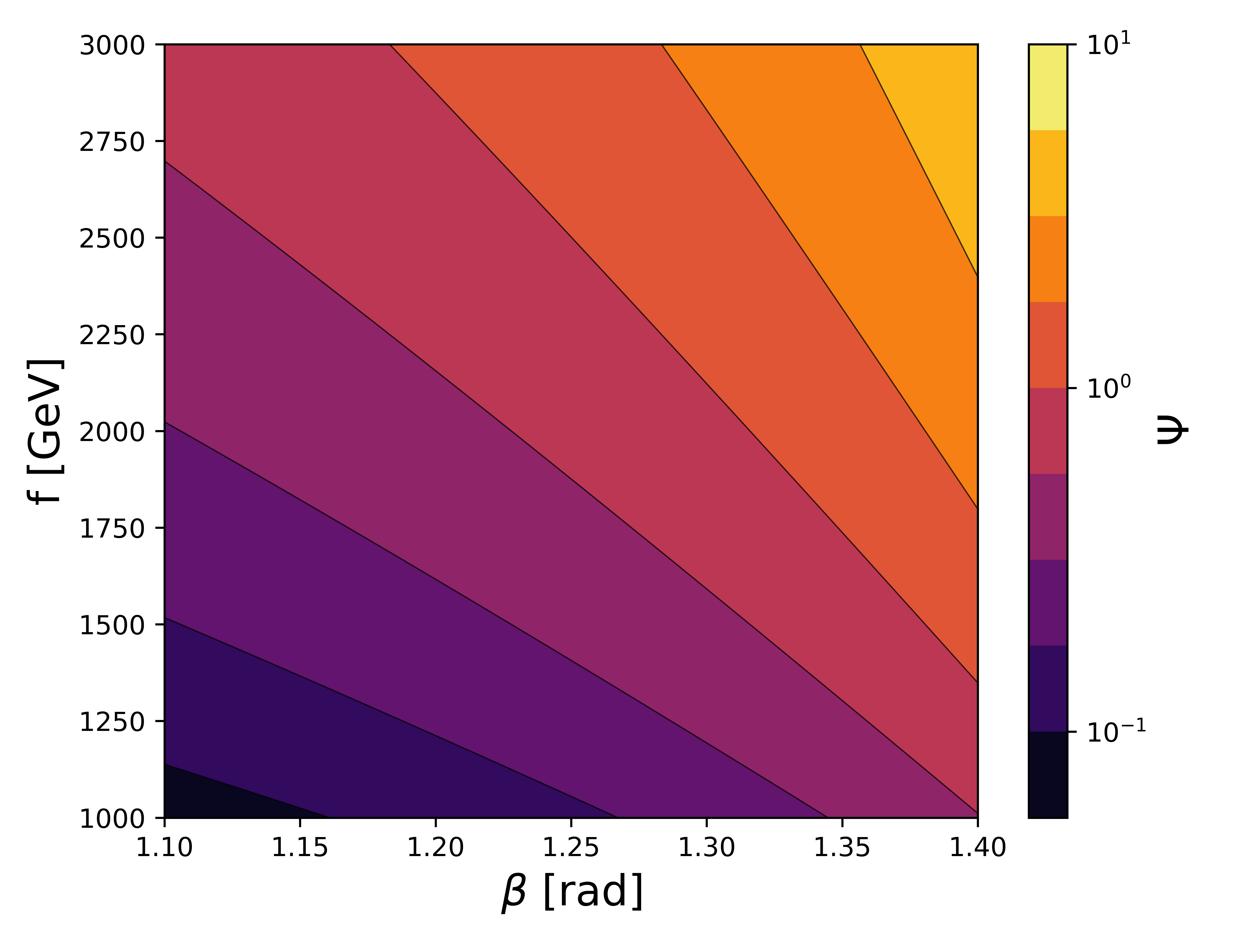}
  }
  \subfigure[]{
    \includegraphics[width=0.485\textwidth]{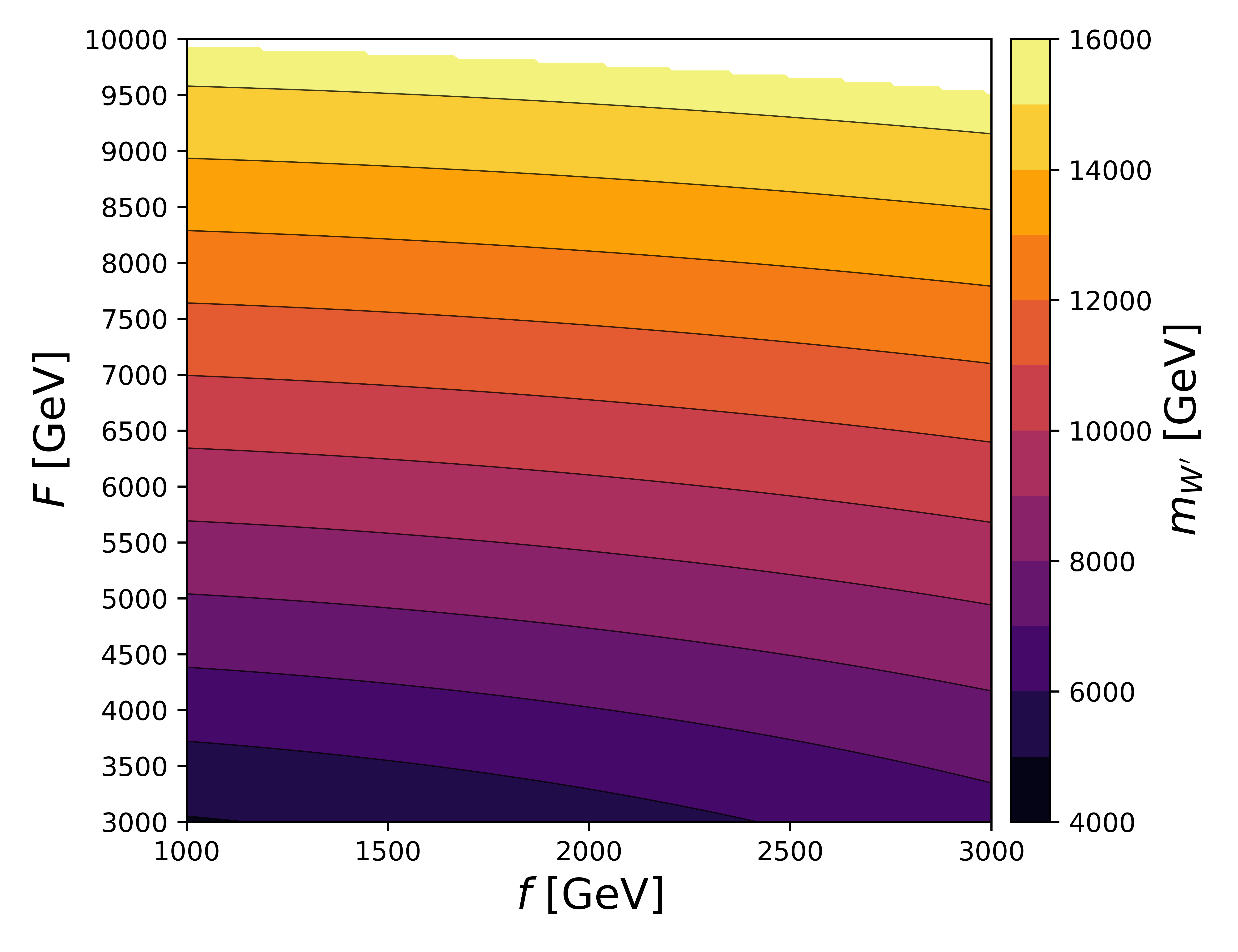}
  }
  \caption{Contour plots of the dependence of some selected BLHM parameters on scalar boson masses, the mixing angle $\beta$, and the symmetry-breaking scales $f$ and $F$. (a) The bilinear coupling $B_{\mu}$ as a function of the pseudoscalar mass $m_{A^0}$ and $\beta$. (b) $B_{\mu}$ as a function of the charged Higgs mass $m_{H^\pm}$ and $\beta$. (c) Fine-tuning parameter $\Psi$ as a function of $f$ and $\beta$. (d) Charged gauge mass $m_{W'}$ as a function of the two symmetry-breaking scales $f$ and $F$. }
  \label{fig:tres_graficas}
\end{figure*}


All of these constraints restrict the BLHM parameters to the ranges shown in Table~\ref{masas-esc1} and Fig.~\ref{fig:tres_graficas}. Table~\ref{masas-esc1} provides a comprehensive listing of parameters and particle masses in the BLHM, constrained at different energy scales: 1 TeV, 2 TeV, and 3 TeV. Additionally, we present parameter values for the limits of the allowed $\beta$ values. One can observe that $\alpha$ and $y_3$ remain constant across energy scales, while $\Psi$, $B_{\mu}$, and the particle masses vary with $f$. Moreover, $\alpha$, $y_3$, $\Psi$, $B_{\mu}$, $m_{A^0}$, $m_{H^0}$, and $m_{H^{\pm}}$ exhibit significant dependence on $\beta$, whereas the remaining parameters show no such dependence.

In Fig.\ref{fig:tres_graficas}.(a), the mass-like parameter $B_{\mu}$ is plotted as a function of the mass of the neutral scalar boson $A^0$, the charged scalar $H^{\pm}$ and $\beta$. Larger $m_{A^0, H^{\pm}}$ and smaller $\beta$ yield higher $B_{\mu}$ values, with variations of $ 1.25 \times 10^5~\mathrm{GeV}^2$ to $3.75 \times 10^5~\mathrm{GeV}^2$. In Fig.\ref{fig:tres_graficas}.(b) $B_{\mu}$ is plotted as a function of the mass of the scalar boson $H^0$ and $\beta$, showing a similar monotonic decrease with $\beta$ as in (a), but over a narrower $m_{H^0}$ range ($860$–$940~\mathrm{GeV}$). In Fig.\ref{fig:tres_graficas}.(c) the fine-tuning parameter $\Psi$ is shown as a function of $f$ and $\beta$. $\Psi$ increases with both $f$ and $\beta$, ranging from $10^{-1}$ to $10^{1}$, indicating regions of no-tuning to moderate tuning. In Fig.\ref{fig:tres_graficas}.(d) the charged gauge mass $m_{W'}$ is plotted as a function $f$ and $F$. The mass increases predominantly with $F$ (we use $F=5$ TeV) from $\sim 4~\mathrm{TeV}$ to $\sim 16~\mathrm{TeV}$, while the dependence on $f$ is weaker. The limits for the plots in Fig.~\ref{fig:tres_graficas} are in accordance with Table~\ref{masas-esc1}.

{
Experimentally, searches for heavy neutral scalars such as $A^0$ and $H^0$ are consistent with the BLHM mass ranges. ATLAS analyzed $A \to ZH$ decays with $m_{A^0} \in [230, 800]$~GeV and $m_{H^0} \in [130, 700]$~GeV using 139~fb$^{-1}$ at $\sqrt{s} = 13$~TeV~\cite{ATLAS:2020gxx}, while $A \to Zh$ excludes \mbox{$m_{A^0} < 1$~TeV} at 95\% C.L.~\cite{ATLAS:2015kpj}. CMS results also exclude $m_{A^0} < 1$~TeV~\cite{CMS:2019ogx}. Simulated studies at the ILC with 500~fb$^{-1}$ show sensitivity to $m_{A^0} \in [200, 250]$~GeV and $m_{H^0} \in [150, 250]$~GeV~\cite{hashemi2019search}.

Charged Higgs boson searches consider $H^{\pm} \to HW^{\pm}$ with $m_{H^{\pm}} \in [300, 700]$~GeV~\cite{cms2022search} and $H^{+} \to t\bar{b}$ with $m_{H^{+}} \in [200, 2000]$~GeV~\cite{ATLAS:2021upq}. For BLHM vector bosons, $W'$ masses are constrained to 2.2--4.8~TeV~\cite{CMS:2022ncp}, and $Z'$ masses are excluded below 4.7~TeV~\cite{CMS:2021klu}, with other bounds in the 800--3700~GeV range~\cite{CMS:2021fyk}.

In the realm of heavy quarks, the decay $T\to Ht$ or $T\to Zt$ is analyzed in \cite{ATLAS:2023pja}. This study explores proton-proton collisions at $\sqrt{s}=13$ TeV with an integrated luminosity of 139 fb$^{-1}$ at ATLAS, revealing no significant signals at the 95\% C.L. for the mass of the $T$ in the range of $1.6-2.3$ TeV. Similar searches for $T$ and $B$ can be found in \cite{ATLAS:2022tla,ATLAS:2018cye,CMS:2018zkf}. In \cite{ATLAS:2022tla}, they also analyze the possibility of a quark with charge $5/3$ like $T^{5/3}$ decaying to $Wt$, imposing a lower limit for $m_{T^{5/3}}$ of $1.42$ TeV.
}

\section{Calculation of the CMDM}
\label{pheno}

\subsection{Flavor mixing scenarios}
 We consider six cases for the extended CKM matrices in the BLHM for constructing the extended CKM-matrix $V_{Hd}$ \cite{hubisz2006flavor,blanke2006particle}:

\noindent\textbf{Case I.} $V_{Hu}=\mathbf{1}$, implying $V_{Hd}=V_{CKM}^{\dagger}$,  represents a benchmark case scenario chosen to provide a basis for comparison. This choice conveys the absence of new sources of CP violation beyond those already contained in the SM, thus weakly receiving contributions from the kind we will address in the cases below.

\noindent\textbf{Case II.} $V_{Hd}=\mathbf{1}$, this implies $V_{Hu}=V_{CKM}^{\dagger}$. Similar to Case 1, this matrix serves as a benchmark scenario where only SM sources of CP violation are present, however, in this case the dominant contribution to the CMDM is expected to arise from mixing with the SM bottom quark.

\noindent\textbf{Case III.} $s_{23}^d=1/\sqrt{2}$, $s_{12}^d=s_{13}^d=0$, $\delta_{12}^d=\delta_{23}^d=\delta_{13}^d=0$. This choice, which departs from the standard CKM matrix, is expected to allow the inclusion of additional effects arising from mixing within the BLHM. The structure of the $V_{Hd}$ matrix implies that the first generation of the heavy quark sector does not mix, while maximal mixing is introduced between the second and third generations by setting $\theta_{23} = \pi/4$.  In this scenario, all phases are set to zero in order to suppress CP violation originating from BSM sources. Substituting the values of Case III into the matrix $V_{Hd}$ in Eq.~(\ref{VHd}), we obtain the matrix:

\begin{equation}
 V_{Hd}=
 \begin{pmatrix}
  1 & 0 & 0\\
  0 & 1/\sqrt{2} & 1/\sqrt{2}\\
  0 & -1/\sqrt{2} & 1/\sqrt{2}
 \end{pmatrix},
 \end{equation}
\\
 \noindent and through the product $(V_{Hd}^{\dagger})^{-1}V_{CKM}^{\dagger}$, we obtain the matrix $V_{Hu}$ =

\begin{equation}
\label{caso3mat}
\begin{aligned}
& \left(
\begin{matrix}
0.9737 \pm 0.0003 & 0.2210 \pm 0.0040 & 0.0080 \pm 0.0003 \\
0.1613 \pm 0.0005 & 0.7183 \pm 0.0043 & 0.7460 \pm 0.0205 \\
-0.1559 \pm 0.0005 & -0.6606 \pm 0.0043 & 0.6873 \pm 0.0205
\end{matrix}
\right)
\end{aligned}
\end{equation}
where the uncertainties are computed analytically using the reported experimental uncertainties of the CKM matrix elements. The latter are found in Table~\ref{tab:exp_val}.\\

\noindent\textbf{Case IV.} $s_{23}^d=0.04183$, $s_{12}^d=0.22501$, $s_{13}^d=0.5$, $\delta_{12}^d=\delta_{23}^d=0$, $\delta_{13}^d=1.147$. In this case, the value $s_{13}^d = 0.5$ is chosen, which is significantly larger than its SM counterpart, with the intention of enhancing mixing between the first and third generations. Additionally, $s_{23}^d = 0.04183$ corresponds to $|V_{cb}|\approx0.041 $, while $ \delta_{13}^d = 1.147 $ matches the CKM phase. This setup is designed to capture both SM and BSM effects within the same framework. Substituting the values of case IV into the matrix $V_{Hd}$ in Eq.~(\ref{VHd}), we obtain the matrix $V_{Hd}$:
\begin{equation}
 \begin{pmatrix}
 0.8438 & 0.1948 & 0.2056-0.4557i\\
  -0.2331-0.0185i & 0.9715-0.0042i & 0.03622\\
  -0.1907-0.4436i & -0.0869-0.1024i & 0.8652
 \end{pmatrix},
 \end{equation}
and $V_{Hu}$:

\begin{widetext}
\begin{equation}
\label{caso4mat}
\begin{aligned}
& \left(
\begin{matrix}
0.8661 \pm 0.0003 + (0.0017 \pm 0.0001)i & 0.3849\pm 0.0036 + (0.0186 \pm 0.0006)i & 0.2233 \pm 0.0059 + (0.4622 \pm 0.0132)i \\
-0.0091 \pm 0.0007 + (0.0191 \pm 0.0001)i & 0.8972\pm 0.0059 + (0.0083 \pm 0.0001)i & 0.0752 \pm 0.0014 + (0.0003 \pm 0.0001)i \\
-0.2019 \pm 0.0002 + (0.4550 \pm 0.0002)i & -0.0916 \pm 0.0015 + (0.1979 \pm 0.0019)i & 0.8723\pm 0.0250 + (0.0078 \pm 0.0002)i
\end{matrix}
\right).
\end{aligned}
\end{equation}
\end{widetext}
\noindent\textbf{Case V.} $s_{23}^d=0.04183$, $s_{12}^d=0.22501$, $s_{13}^d=0.5$, $\delta_{12}^d=\delta_{23}^d=\delta_{13}^d=0$. Similar to Case IV, but with all the phases set to zero which means that all the CP violation will come from the SM mixing. Substituting the values of case V into the matrix $V_{Hd}$ in Eq.~(\ref{VHd}), we obtain the matrix $V_{Hd}$:
\begin{equation}
 \begin{pmatrix}
  0.8438 & 0.1948 & 0.5\\
  -0.2451 & 0.9687 & 0.03622\\
  -0.4773 & -0.1531 & 0.8652
 \end{pmatrix},
 \end{equation}
and $V_{Hu}$:
\begin{equation}
\label{caso5mat}
\begin{aligned}
& \left(
\begin{matrix}
0.8672\pm0.0003  & 0.3969\pm 0.0037 & 0.5222\pm 0.0145 \\
-0.0213\pm0.0008 & 0.8919\pm 0.0008 & 0.0750\pm 0.0014 \\
-0.4958 \pm0.0003 & -0.2196\pm0.0025 & 0.8677\pm 0.0252
\end{matrix}
\right).
\end{aligned}
\end{equation}
\noindent\textbf{Case VI.} $s_{23}^d=0.4$, $s_{12}^d=0.7$, $s_{13}^d=0.5$, $\delta_{12}^d=\delta_{23}^d=\delta_{13}^d=0$. Similarly, setting all phases to zero implies that CP violation from BSM sources is not expected in this scenario. 
Moreover, the choice $ s_{12}^d = 0.7$ corresponds to a large mixing angle compared to the Cabibbo angle. Likewise, $s_{23}^d = 0.4$ and $s_{13}^d = 0.5$ differ significantly from their SM counterparts, with the latter introducing flavor mixing that is absent in the SM. This scenario is constructed to highly depart from SM behavior and aims to capture potential effects arising from BSM physics. Substituting the values of case VI into the matrix $V_{Hd}$ in Eq.~(\ref{VHd}), we obtain the matrix $V_{Hd}$:
\begin{equation}
 \begin{pmatrix}
  0.6184 & 0.6062 & 0.5\\
  -0.7843 & 0.5144 & 0.3464\\
  -0.0472 & -0.6064 & 0.7936
 \end{pmatrix},
 \end{equation}
and $V_{Hu}$:
\begin{equation}
\label{caso6mat}
\begin{aligned}
& \left(
\begin{matrix}
0.7401\pm 0.0001 & 0.7481\pm0.0045 & 0.5371\pm0.0145 \\
-0.6471\pm 0.0005 & 0.3424\pm0.0044 & 0.3663\pm0.0101 \\
-0.1790\pm0.0005 & -0.5693\pm0.0038 & 0.779\pm0.0229
\end{matrix}
\right).
\end{aligned}
\end{equation}

\subsection{CMDM calculation and error propagation}

To compute the amplitudes in Eq.~(\ref{ampE})-(\ref{ampV}), we use the \texttt{FeynCalc} package \cite{shtabovenko2020feyncalc} and the \texttt{Package X} \cite{patel2015package} for \texttt{Mathematica}. We obtain the total contribution of the BLHM to the chromomagnetic dipole by summing the scalar and vector contributions given by the amplitudes of Eqs.~(\ref{ampE}) and (\ref{ampV}). The computation of $\hat{\mu}_t^{\text{BLHM}}$ involves several measured boson and quark masses, as well as Standard Model (SM) parameters, whose values are listed in Table~\ref{tab:exp_val}. These values are taken from the Particle Data Group \cite{Workman:2022ynf} and include both statistical and systematic uncertainties, hence becomes essential to propagate them into our calculations. To achieve this, we perform a statistical simulation for error propagation: We randomly sample the experimental masses and the SM parameters from a Gaussian probability distribution with a
mean equal to their central experimental value and a width equal to
the squared sum of the uncertainties. We calculate $\hat{\mu}_t^{BLHM}$ using sampled masses and parameters corresponding to the experimentally observed SM masses parameters, and the given values for the BLHM parameters. We repeat the
procedure $10^4$ times. In this manner, we obtain a
Gaussian distribution for the $\hat{\mu}_t^{BLHM}$. Next, we
assign the mean of the distribution as the value of the
parameter and used its difference from the distribution
quantiles at 68\% and 95\% confidence level (C.L.), in order to extract the 68\% and 95\% confidence intervals (C.I.). {\color{black} The latter allows for asymmetric uncertainties}. This method is known as the Monte Carlo
bootstrap uncertainty propagation \cite{Efron1994, Molina:2020zao}. 

\begin{table}[h!]
\caption{Masses and parameters with uncertainties used  in the calculation of the $\hat{\mu}_t^{BLHM}$.}
\begin{tabular}{c | c }\hline \hline
      Parameter           & Value  \\ \hline
$m_{Z}$       & $91.1876 \pm 0.0021$ GeV 
\\
$m_{W}$          & $80.377 \pm 0.012$ GeV
\\
$m_{h^0}$       & $125.11 \pm 0.11$  GeV 
\\
$m_{t}$       & $172.76 \pm 0.3$  GeV 
\\
$m_{b}$       & $ 4.18\pm 0.00045$  GeV
\\
$g$       & $ 0.6258\pm 0.001$ 
\\
$g'$       & $ 0.3528\pm 0.001$ 
\\
$|V_{ud}|$       & $ 0.97373\pm 0.00031$ 
\\
$|V_{us}|$       & $ 0.2243\pm 0.0008$ 
\\
$|V_{ub}|$       & $ 0.00382\pm 0.00020$ 
\\
$|V_{cd}|$       & $ 0.221\pm 0.004$ 
\\
$|V_{cs}|$       & $ 0.975\pm 0.006$ 
\\
$|V_{cb}|$       & $ 0.0408\pm 0.0014$ 
\\
$|V_{td}|$ & $0.0086\pm 0.0002$ 
\\
$|V_{ts}|$ & $0.0415\pm 0.0009$ 
\\
$|V_{tb}|$ & $1.014\pm 0.029$ 
\\
$v$ & $246.22\pm (6.06\times 10^{-12})$ GeV 
\\


\hline\hline
\end{tabular}
\label{tab:exp_val}
\end{table}

\section{Results}
\label{results}
This section presents our results for the $\hat{\mu}_t^{BLHM}$ calculations. \mbox{Figs.~\ref{caso1_allbetas}-\ref{TodosCasos_TodosBetas}} condense the results for all the studied $\beta$ angles for Cases I-VI, as described in section~\ref{pheno}. Additionally, we provide an individual plot for each $\beta$ considered, to show the 68\% and 95\% confidence interval bands arising from propagating the experimental and model parameter uncertainties. These plots are shown in Fig.~\ref{caso1Indi}, and correspond to Cases I-VI. In Table~\ref{CMDMS}, we summarize the numerical values of our $\hat{\mu}_t^{BLHM}$ calculations along with their uncertainty at 68\% C.L.

\subsection{\textbf{Parameter search}}
\label{params}

\begin{figure*}[htbp]
\centering
\subfigure[]{
\includegraphics[width=8.5cm]{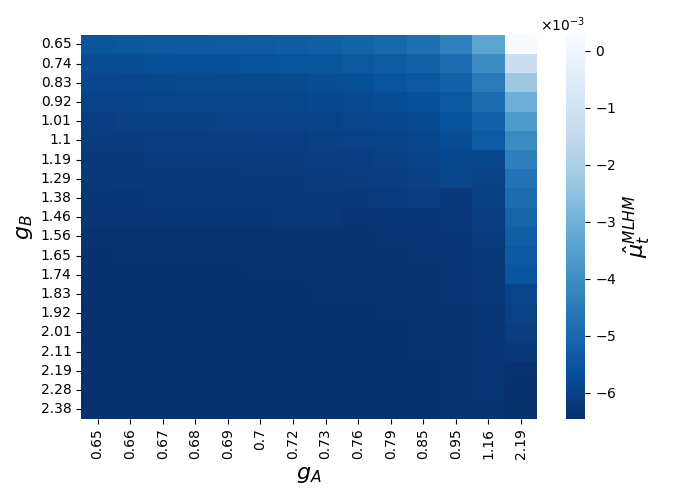}
}
\subfigure[]{
\includegraphics[width=8.5cm]{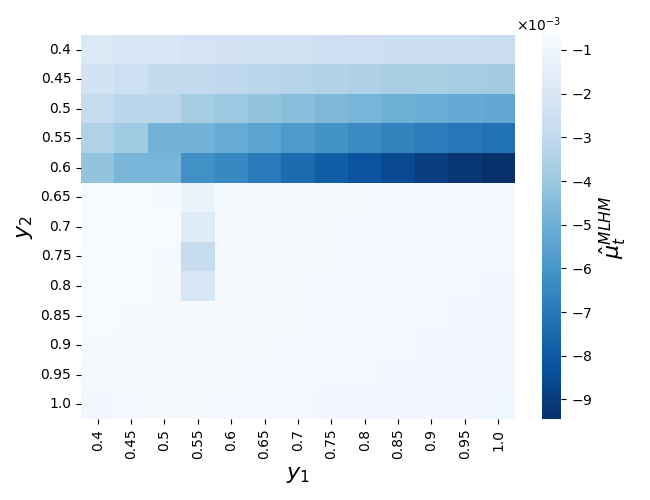}
}

\caption{ (a) Dependence of $\hat{\mu}_t^{\text{BLHM}}$ on the parameters $g_A$ and $g_B$. (b) CMDM dependence on the ratio between $y_1$ and $y_2$.}
\label{fig:params}
\end{figure*}

In addition to the BLHM parameter space discussed in Sec.~\ref{pspace}, we examine, given their theoretical relevance, the dependence of $\hat{\mu}_t^{\text{BLHM}}$ on the gauge couplings $g_A$ and $g_B$ in Eq.~(\ref{acoplesSU})-(\ref{acoplesSU1}), and the Yukawa couplings $y_1$, $y_2$, and $y_3$, as defined in Eqs.~(\ref{masa-top})–(\ref{acople-yt}). The dependence of $\hat{\mu}_t^{\text{BLHM}}$ on $g_A$ and $g_B$ is shown in Fig.\ref{fig:params}a. Note that the variation of $\hat{\mu}_t^{\text{BLHM}}$ is below 1\% for most of the cases. This can be explained by the fact that $g_A$ and $g_B$ are expected to play similar roles due to their symmetric appearance in the BLHM. Figure~\ref{fig:params}b shows the dependence of $\hat{\mu}_t^{\text{BLHM}}$ on the Yukawa couplings. In this case, the variations are more pronounced: the maximum value of the chromomagnetic moment found in this study is approximately one order of magnitude larger than the minimum value. This behavior can be attributed to the fact that the Yukawa couplings enter the calculations more directly, as they appear, for example, in all fermion masses.
Along this line, we choose optimal values for the aforementioned parameters, such that the masses of the BLHM quarks and bosons are in range of the experimental allowed values: $g_A=2.1,\; g_B =2.3,\; y_1=0.5$, and $ y_2=0.9$. The following calculations are performed using this parameter set.

\subsection{\textbf{Cases I-VI}}
\label{Scase1}

Figure~\ref{caso1_allbetas} presents the CMDM for the first case of the extended CKM matrix, with $V_{Hu}=\mathbf{1}$ and \mbox{$1<f<3$ TeV}. The figure also shows the results for the $\beta=1.10,1.20,1.30,1.40$ angles. Note that the magnitude of $\hat{\mu}_t$ remains in the order of $10^{-3}$, independently of the value of $\beta$, although a dependence is clearly observed. This behavior can also be seen in the rest of the cases, as will be shown.

\begin{figure}
\centering
\includegraphics[width=8.5cm]{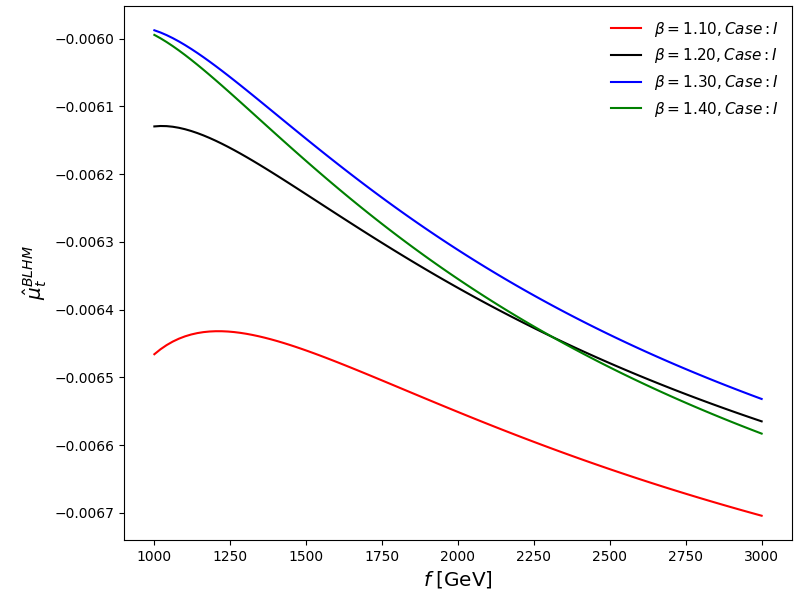}
\caption{\textbf{Case I.} Total contributions to the CMDMs with the CKM matrix $V_{Hu}=\mathbf{1}$ and four different $\beta$ angles. }
\label{caso1_allbetas}
\end{figure}

\begin{figure}
\centering
\includegraphics[width=8.5cm]{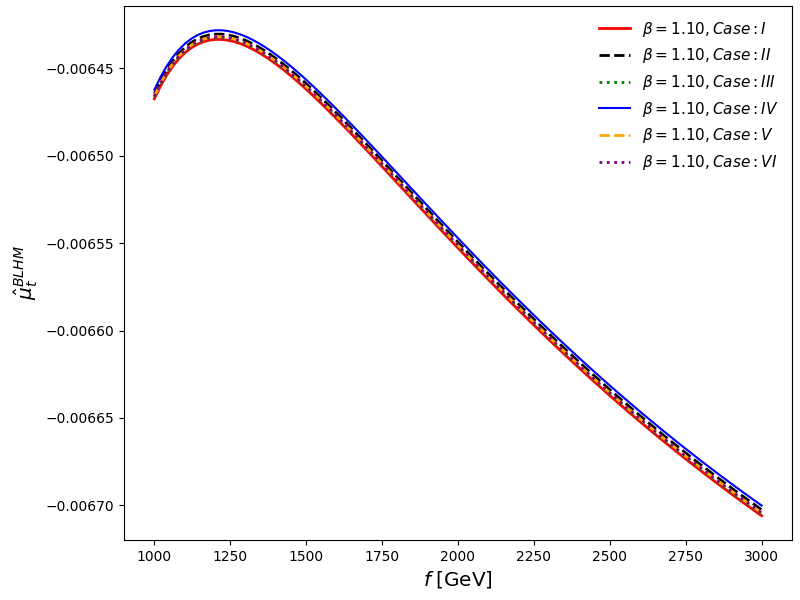}
\caption{\textbf{Cases I-VI.} Total contributions to the CMDMs with the full spectrum of possible CKM matrix ($V_{Hu}$) values, for a $\beta$ value of 1.10 rad.}
\label{casos1a6}
\end{figure}

In Fig. \ref{casos1a6}, we plot all six cases for \mbox{$\beta=1.10$ rad.} We note the formation of two groups, one with two curves, while the other with four nearly overlapping curves, corresponding from top to bottom, to \mbox{Cases II, IV} and \mbox{Cases I, III, V, VI}. This allows us to treat the current CMDM results as though only two cases effectively exist, namely, \mbox{Case II, IV} and \mbox{Case I, III, V, VI}. Profiting from this, Fig. \ref{TodosCasos_TodosBetas} plots all cases for all $\beta$'s, where the solid lines represent \mbox{Case I, III, V, VI}, while the dashed ones represent \mbox{Case II, IV}. Notice that the plot encompass all I-VI studied Cases, via displaying four pairs of curves, the blue, green, black and red ones, corresponding from top to bottom to the \mbox{$\beta=1.30, 1.40, 1.20$ and $1.10$ rad} angles, respectively. Furthermore, note that the Case splitting behavior seen in \mbox{Fig. \ref{casos1a6}} for the $\beta$ value of $1.10$, now holds for the rest of the $\beta$ values, but in a minor degree — practically vanishing for \mbox{$\beta=1.20$}. Although we expected to observe a stronger separation between the six different scenarios, the similarity can be attributed to the fact that, in this work, we only consider flavor-changing interactions between the SM top quark and the SM $b$ quark, as well as the BLHM heavy $B$ quark. As seen in Eq.(\ref{caso3mat})–(\ref{caso6mat}), the matrix elements, which encode the information of these flavor-changing processes, are quite close to each other. Despite that the difference between them can be as large as  30\% it does not significantly propagate through the final calculation of the top quark's CMDM. This result opens the opportunity to explore the possibility of mixing with other quark generations in future investigations, where the elements of different extended matrices may have a greater impact. In order to visualize the impact of experimental uncertainties for our effective groups, we present individual plots of the $\hat{\mu}_t^{BLHM}$ for each of them, for the different $\beta$ angles. These plots are intended to ease the display of the confidence intervals in our calculations, due to error propagation. Fig.~\ref{caso1Indi} shows the $\hat{\mu}_t^{BHLM}$ for Cases \mbox{I, III, V, VI} (upper plots) and \mbox{II, IV} (lower plots), as discussed in Section~\ref{Scase1}. 
Note that in both effective cases, the error bands resulting from experimental uncertainties yield values on the order of $10^{-3}$. This indicates that our calculations are stable and consistent. Moreover, the main variations in the central values of our calculations for $\hat\mu_t^{\text{BLHM}}$ arise primarily from the variation of the model parameters: $\beta$, $y_1$, $y_2$, $y_3$, and $f$. Thus, these error bands may help determine whether a deviation in the measured CMDM is truly a prediction of the model or merely a consequence of experimental uncertainties.

\begin{figure}
\centering
\includegraphics[width=8.5cm]{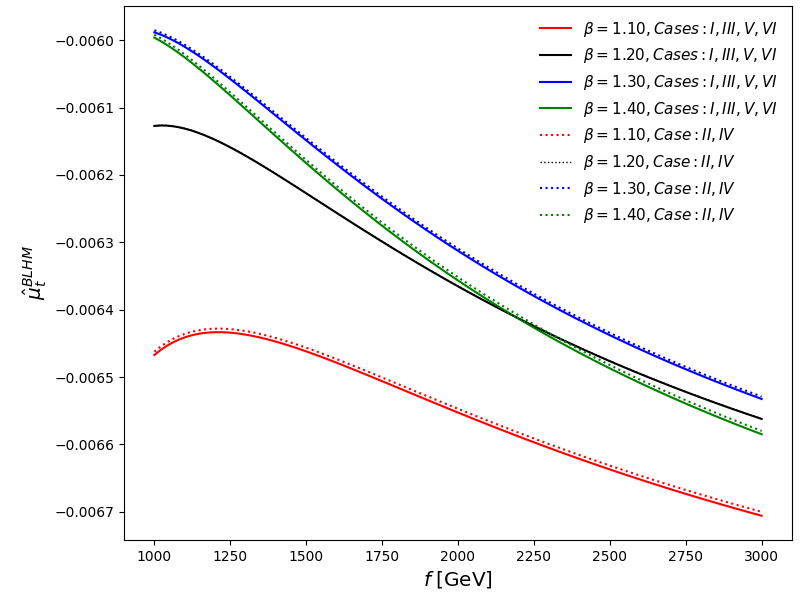}
\caption{\textbf{Cases I-VI.} Total contributions to the CMDMs with the full spectrum of possible CKM matrix ($V_{Hu}$) and $\beta$ values. The solid lines represent Case I, III, V, VI, while the dashed ones represent Case II, IV.}
\label{TodosCasos_TodosBetas}
\end{figure}

Finally, Table~\ref{CMDMS} displays our numerical values for the CMDM according to $f=1,3$ TeV. 
{\color{black} As suggested by our plots, these values are nearly identical across all six Cases.}


\begin{figure*}[htbp]
  \subfigure[]{
    \includegraphics[width=0.35\textwidth]{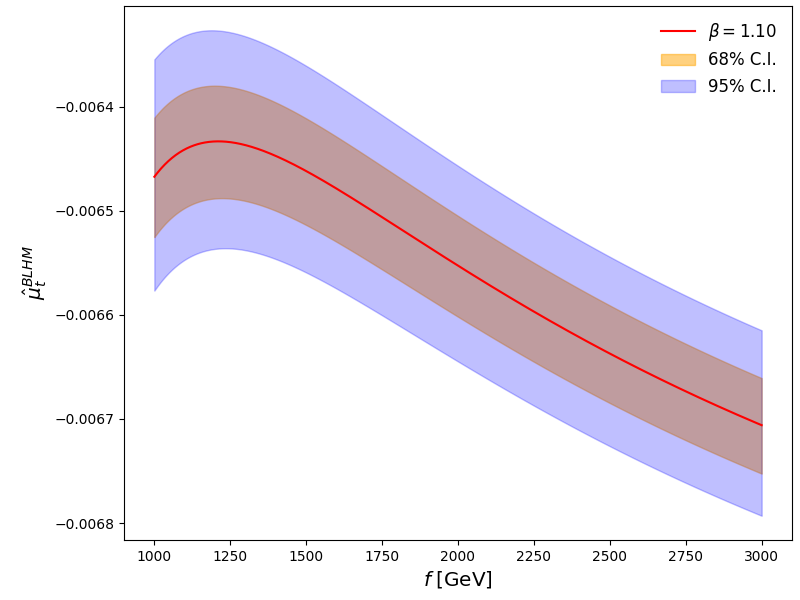}
  }    
  \subfigure[]{
    \includegraphics[width=0.35\textwidth]{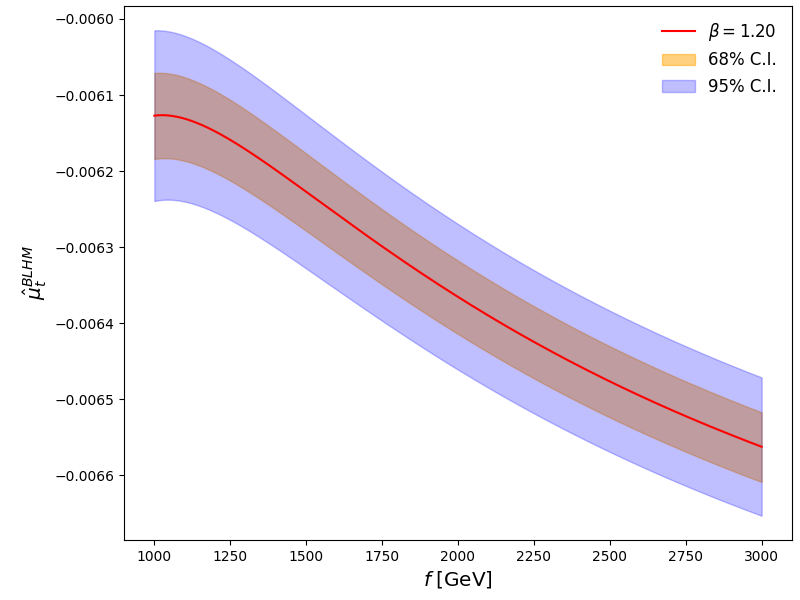}
  }
  \subfigure[]{
    \includegraphics[width=0.35\textwidth]{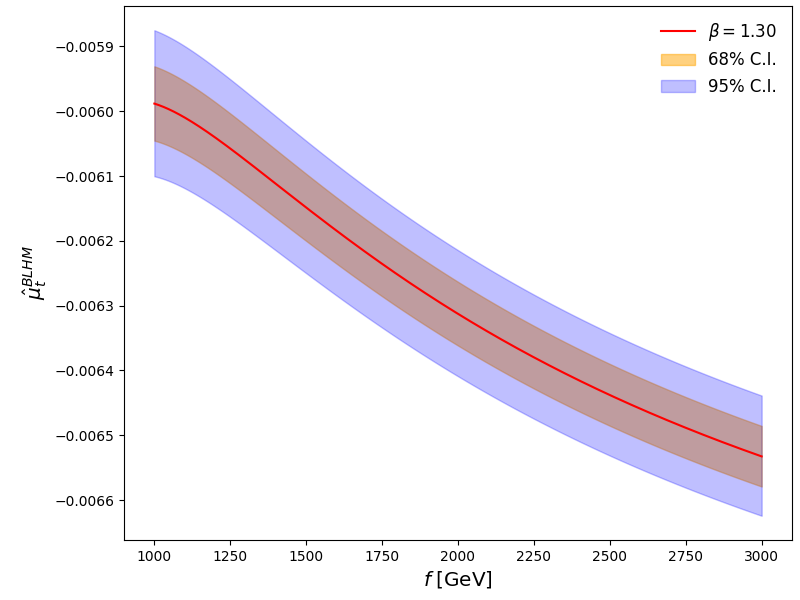}
  }
  \subfigure[]{
    \includegraphics[width=0.35\textwidth]{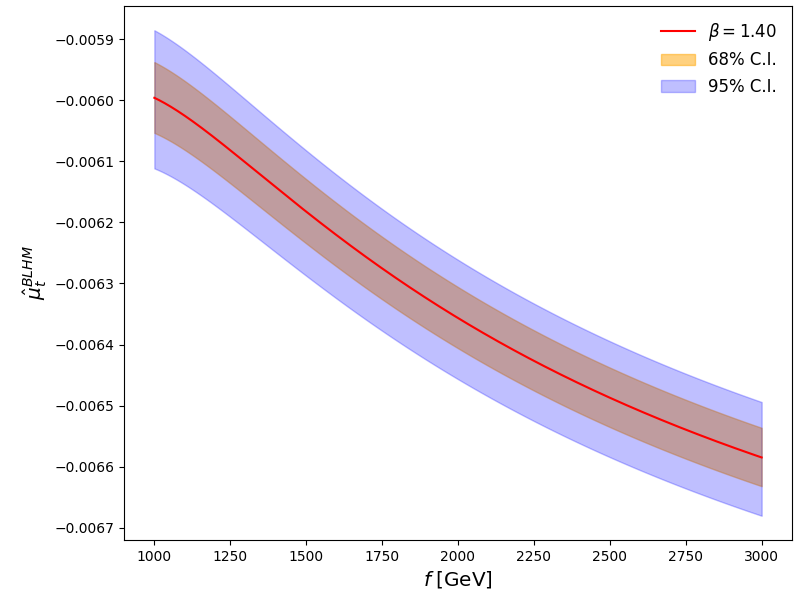}
  }
  \subfigure[]{
    \includegraphics[width=0.35\textwidth]{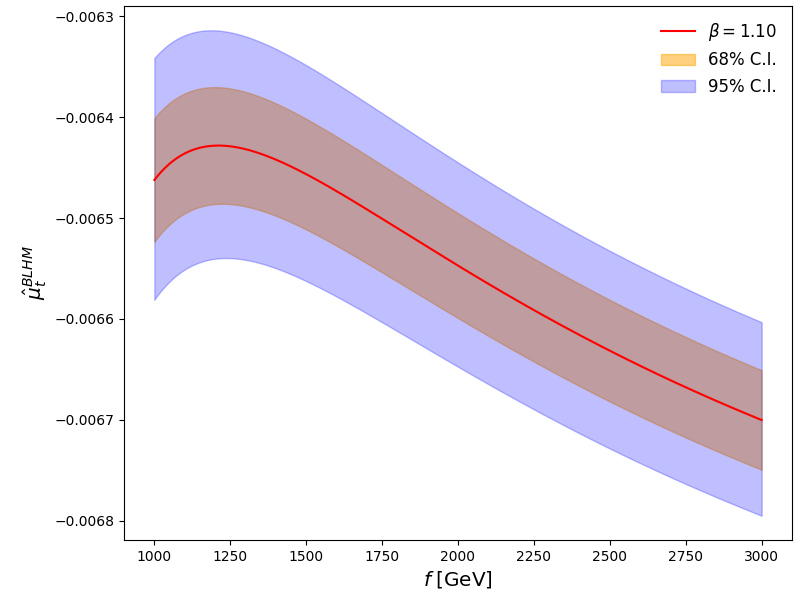}
  }
  \subfigure[]{
    \includegraphics[width=0.35\textwidth]{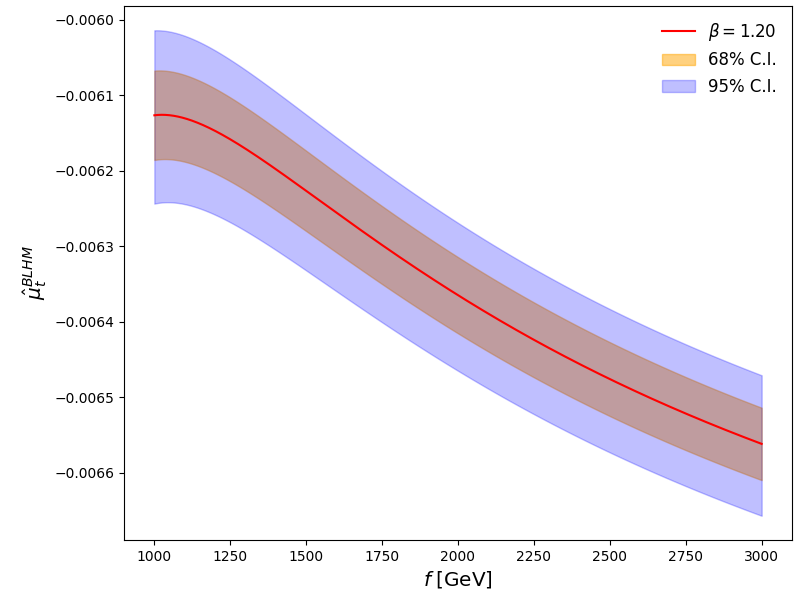}
  }
  \subfigure[]{
    \includegraphics[width=0.35\textwidth]{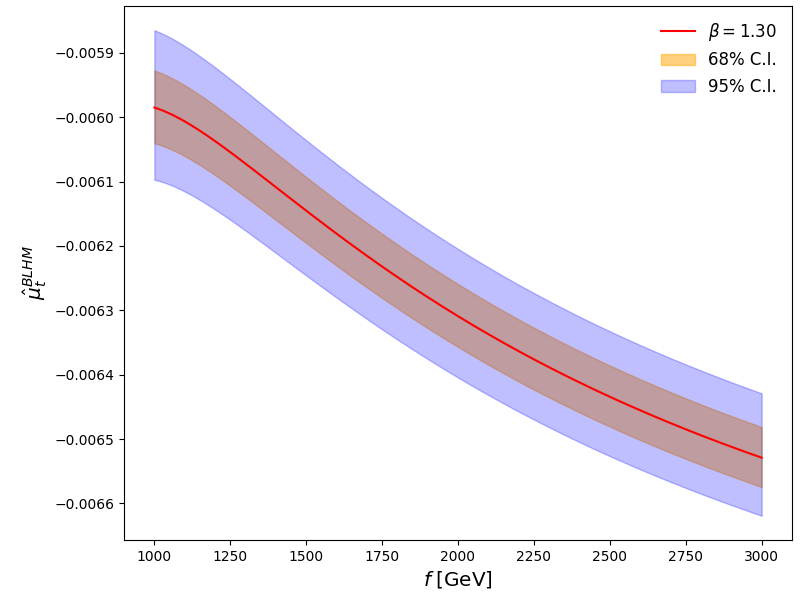}
  }
  \subfigure[]{
    \includegraphics[width=0.35\textwidth]{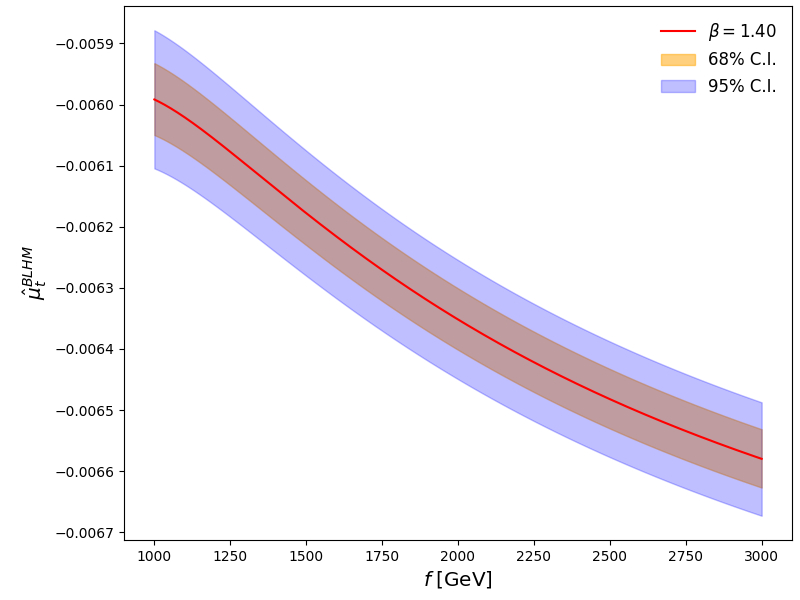}
  }
  \caption{\textbf{Cases I, III, V, VI and II, IV .} {\color{black}Total contributions to the CMDM for $\beta$ angle indicated in each plot. Plots (a)-(d) refer to Case \mbox{I, III, V, VI}, while (e)-(h) plots are for Case \mbox{II, IV}. The solid reds lines correspond to the central values of our calculations. The orange and blue bands are the 68\% and 95\% C.I., respectively, which arise from the uncertainty propagation described in Sec.\ref{pheno}}.}
  \label{caso1Indi}
\end{figure*}

\begin{table*}
\caption {Numerical values of the CMDM for cases I,II,V,VI and II-IV of the extended CKM matrix. {\color{black} In the first column the $\beta$ angles are indicated. The second, third, and fourth columns contain the computed CMDM values for $f=1, 2, 3$, TeV, respectively. The uncertainties are reported at 68\% C.L. }} \label{CMDMS}
\medskip
\resizebox{\columnwidth}{!}{%
\begin{tabular}{c| c c c }\hline \hline

& & Cases I,II,V,VI &\\
\hline
$ \beta $ [rad] & $f=1$ TeV & $f=2$ TeV & $f=3$ TeV \\ \hline 
$1.1$ & $(-6.47^{+0.06}_{-0.06})\times 10^{-3}$ & $(-6.55^{+0.05}_{-0.05})\times 10^{-3}$ & $(-6.71^{+0.05}_{-0.05})\times 10^{-3}$ \\ 
$1.2$ & $(-6.13^{+0.06}_{-0.06})\times 10^{-3}$ & $(-6.37^{+0.05}_{-0.05})\times 10^{-3}$ & $(-6.56^{+0.05}_{-0.05})\times 10^{-3}$ \\ 
1.3 & $(-5.99^{+0.06}_{-0.06})\times 10^{-3}$ & $(-6.31^{+0.05}_{-0.05})\times 10^{-3}$ & $(-6.53^{+0.05}_{-0.05})\times 10^{-3}$ \\ 
1.4 & $(-6.0^{+0.06}_{-0.06})\times 10^{-3}$ & $(-6.36^{+0.05}_{-0.05})\times 10^{-3}$ & $(-6.58^{+0.05}_{-0.05})\times 10^{-3}$ \\ 
\hline \hline

& & Cases II,IV &\\
\hline
$ \beta $ [rad] & $f=1$ TeV & $f=2$ TeV & $f=3$ TeV \\ \hline 
$1.1$ & $(-6.46^{+0.06}_{-0.06})\times 10^{-3}$ & $(-6.55^{+0.05}_{-0.05})\times 10^{-3}$ & $(-6.7^{+0.05}_{-0.05})\times 10^{-3}$ \\ 
$1.2$ & $(-6.13^{+0.06}_{-0.06})\times 10^{-3}$ & $(-6.37^{+0.05}_{-0.05})\times 10^{-3}$ & $(-6.57^{+0.05}_{-0.05})\times 10^{-3}$ \\ 
1.3 & $(-5.99^{+0.06}_{-0.06})\times 10^{-3}$ & $(-6.31^{+0.05}_{-0.05})\times 10^{-3}$ & $(-6.53^{+0.05}_{-0.05})\times 10^{-3}$ \\ 
1.4 & $(-5.99^{+0.06}_{-0.06})\times 10^{-3}$ & $(-6.35^{+0.05}_{-0.05})\times 10^{-3}$ & $(-6.58^{+0.05}_{-0.05})\times 10^{-3}$ \\ 
\hline \hline

\end{tabular}
}

\end{table*}

\subsection{Comparison with other works}

\begin{table*}
\caption{Comparison of the absolute value of the top quark chromomagnetic dipole moment, $|\hat{\mu}_t|$, reported in various studies. SM predictions  are displayed in the upper block. BSM predictions are displayed in the lower block. This work's result, for the flavor enhanced BLHM $|\hat{\mu}_t|$, is found and highlighted in the lower block.
}
\begin{tabular}{c | c}
\hline
\hline

Model & $\vert \hat{\mu}_{t} \vert$ \\
\hline
SM  & $10^{-2}$~\cite{Martinez:2001qs}\\
SM  & $10^{-2}$~\cite{Martinez:2007qf}\\
SM  & $10^{-2}$~\cite{Aranda:2020tox}\\
SM  & $10^{-2}$~\cite{tututi2023}\\

\hline\hline

Enhanced BLHM  & $10^{-3}$ \textbf{[This work]} \\
BLHM  & $10^{-6}-10^{-5}$ ~\cite{Aranda:2021kza} \\
2HDM-II  & $10^{-3}$ ~\cite{Martinez:2007qf}\\
Technicolor TC2  & $10^{-2}$ ~\cite{Martinez:2007qf}\\
Extra dimensions 5D  & $10^{-3}$ ~\cite{Martinez:2007qf}\\
LH T-parity & $10^{-3}$$-10^{-4}$ ~\cite{Ding:2008nh} \\
331  & $10^{-5}$~\cite{Hernandez-Juarez:2020gxp}~\cite{montano1}\\
$Z^{\prime}$ FCNC  & $10^{-6}$~\cite{montano1}\\

Decoupling EFT  & $<10^{-2}$ ~\cite{Martinez:1996cy} \\

4GTHDM  & $10^{-2}$-$10^{-1}$~\cite{Hernandez-Juarez:2018uow} \\
Uniparticle  & $10^{-2}$~\cite{Sampayo:2010}\\

Two-loop 2HDM  & $10^{-4}$-$10^{-3}$ ~\cite{Bisal2024} \\
\hline
\hline
\end{tabular}

\label{CMDMcompare}
\end{table*}

{Table~\ref{CMDMcompare} presents a comparison of the absolute value of the top quark chromomagnetic dipole moment, $|\hat{\mu}_t|$, for SM and BSM calculations, as reported in various references. 
The SM predictions (upper block) consistently yield values of order $10^{-2}$. 
In contrast, BSM models (lower block) exhibit a broader range of predictions: values between $10^{-3}$ and $10^{-2}$ are obtained in the \mbox{2HDM-II,} Technicolor TC2, and 4GTHDM frameworks, while significantly smaller values, down to $10^{-6}$, are predicted in the \mbox{$Z^{\prime}$ FCNC.}
Our result, obtained within the flavor-enhanced BLHM framework, is labeled in the table as ``This work" and gives $|\hat{\mu}_t| \sim 10^{-3}$. 
The wide range of predictions highlights the sensitivity of $\hat{\mu}_t$ to new physics effects, particularly those involving extended scalar, gauge, or fermionic sectors that directly impact the top-quark interactions.

These BSM contributions are expected to modify the SM values, hence any experimentally measured deviation of the CMDM w.r.t. the SM calculations might serve as an indication of BSM physics. The current CMDM experimental sensitivity achieved at the LHC is of the order $10^{-3}$, therefore, enhancing upon the latest BLHM CMDM predictions, all lying below $10^{-4}$, constitutes a relevant improvement. Furthemore, when comparing our results with other BSM predictions, we find that the inclusion of flavor mixing in the BLHM framework makes our calculation competitive with results obtained in other BSM scenarios.}

\section{Conclusions}
\label{conclusions}
In this work, we compute the CMDM of the top quark within the BLHM framework. For the first time, we include contributions from flavor-changing interactions in the BLHM, introduced through an extended CKM matrix. We study the top-quark CMDM across a range of experimentally allowed values of the $\tan\beta$ parameter. Within the allowed BLHM parameter space—constrained by Higgs sector corrections and fine-tuning considerations—we find CMDM values on the order of $10^{-3}$. This represents a significant improvement over previous BLHM results~\cite{Aranda:2021kza}, which did not account for flavor mixing effects.

The CMDM serves as an alternative probe for BSM heavy quarks, particularly given that experimental searches have so far found no evidence for such quarks below 2.3 TeV for $T$ and $B$ \cite{ATLAS:2023pja}, and below 1.42 TeV for exotic states such as $T^{5/3}$, $T^6$, and $T^{2/3} $\cite{ATLAS:2022tla}. However, the current experimental precision on $\hat{\mu}_t$ remains limited, with uncertainties as large as 85\%. Further improvements in measurement precision are therefore necessary for this observable to become competitive with direct searches.

Finally, we have thoroughly propagated all sources of uncertainty to obtain robust uncertainty bands in our CMDM predictions. This enables precise assessment of the extent to which deviations arising from experimental uncertainties can be accommodated within the statistical and theoretical uncertainties of the CMDM BLHM. These results may provide valuable guidance for future theoretical, phenomenological, and experimental investigations at current and future particle colliders. Moreover, the separation observed among our CMDM predictions for the six studied Cases—particularly the smaller differences among \mbox{Cases I, III, V, VI}—can be attributed to the specific flavor-changing processes considered in this work. This outcome highlights the potential for future investigations into mixing with other quark generations.

\begin{acknowledgments}
T. C. P. thanks SECIHTI  postdoctoral fellowship and SNII México. 
\end{acknowledgments}

\appendix

\section{Feynman rules in the BLHM}
\label{feynrules}

In this appendix, we present the Feynman rules that involve flavor-changing in the BLHM. The rest of the vertices used can be found in \cite{Aranda:2021kza,Cruz-Albaro:2023pah}.

Tables \ref{feynrulesTabla}--\ref{feynrulesTabla4} summarize the Feynman rules for the 3-point interactions: fermion-fermion-scalar (FFS), fermion-fermion-gauge (FFV), gauge-gauge-gauge (VVV), and scalar-gauge-gauge (SVV) interactions.

\begin{table}[H]
\centering
\caption{Essential Feynman rules in the BLHM for studying the CMDM.}
\begin{tabular}{|c|c|}\cline{1-2}
\rule{0pt}{4ex} Vertex & Rule (Factors in Tables \ref{feynrulesTabla3} and \ref{feynrulesTabla4})\\\cline{1-2}
\rule{0pt}{4ex} $W^{\prime -}\bar{B}t$ & $\dfrac{-igs_{\beta}v}{4\sqrt{2}f}A_y\gamma^{\mu}P_L(V_{Hu})$ \\\cline{1-2}
\rule{0pt}{4ex} $\eta^{-}\bar{B}t$ & $\dfrac{4im_W^2}{f^2g^2\sqrt{y_1^2+y_2^2}(y_1^2+y_3^2)}(Y_1P_L+Y_2P_R)(V_{Hu})$ \\\cline{1-2}
\rule{0pt}{4ex} $\phi^{-}\bar{B}t$ & $F_aF_b(X_1P_L+X_2P_R)(V_{Hu})$ \\\cline{1-2}
\rule{0pt}{4ex} $H^{-}\bar{B}t$ & $\dfrac{-3\sqrt{2}m_Wc_{\beta}s_{\beta}y_1y_2y_3^2}{fg\sqrt{y_1^2+y_2^2}(y_1^2+y_3^2)}
P_L(V_{Hu})$ \\\cline{1-2}
\end{tabular}
\label{feynrulesTabla}
\end{table}

\begin{table}[H]
\centering
\caption{Factors from Table \ref{feynrulesTabla}.}
\begin{tabular}{|c|c|}\cline{1-2}
\rule{0pt}{4ex} Factor & Expression  \\\cline{1-2}
\rule{0pt}{4ex} $A_y$ & $y_3(2y_1^2-y_2^2)(y_1^2+y_2^2)^2(y_1^2+y_3^2)$ \\\cline{1-2}
\rule{0pt}{4ex} $Y_1$ & $B_1+B_2$ \\\cline{1-2}
\rule{0pt}{4ex} $Y_2$ & $B_3+B_4$ \\\cline{1-2}
\rule{0pt}{4ex} $F_a$ & $\dfrac{y_3(2y_1^2-y_2^2)}{(y_1^2+y_3^2)\sqrt{y_1^2+y_2^2}}$ \\\cline{1-2}
\rule{0pt}{4ex} $F_b$ & $\dfrac{Fs_{\beta}}{2f\sqrt{2(f^2+F^2)}}$ \\\cline{1-2}
\rule{0pt}{4ex} $X_1$ & $A_1+A_3+A_6$ \\\cline{1-2}
\rule{0pt}{4ex} $X_2$ & $A_2+A_4+A_5$ \\\cline{1-2}
\end{tabular}
\label{feynrulesTabla3}
\end{table}

\begin{table}[H]
\centering
\caption{Factors from Table \ref{feynrulesTabla3}.}
\begin{tabular}{|c|c|}\cline{1-2}
\rule{0pt}{4ex} $B_1$ & $y_1y_2(y_1^2+y_3^2)c_{\beta}^2$ \\\cline{1-2}
\rule{0pt}{4ex} $B_2$ & $y_1^3y_2s_{\beta}-\dfrac{1}{2}y_1y_2y_3^2s_{\beta}$ \\\cline{1-2}
\rule{0pt}{4ex} $B_3$ & $\dfrac{fg}{4m_W}y_3(2y_1^2+5y_2^2)\sqrt{y_1^2+y_3^2}s_{\beta}$ \\\cline{1-2}
\rule{0pt}{4ex} $B_4$ & $-y_1^3y_3s_{\beta}-\dfrac{1}{2}y_1(y_2^2+y_3^2)s_{\beta}$ \\\cline{1-2}
\rule{0pt}{4ex} $A_1$ & $\dfrac{-4m_Wc_{\beta}^2y_1y_2(y_1^2+y_3^2)}{s_{\beta}y_3(2y_1^2-y_2^2)}$ \\\cline{1-2}
\rule{0pt}{4ex} \multirow{2}{*}{$A_2$} & $fg-y_1^8-m_Ws_{\beta}y_3y_2^6-3y_2y_1^4(2y_1^2-y_2^2)$ \\
\rule{0pt}{4ex} & $-y_1^6(3y_2^2+y_3^2)-y_1^2y_2^4(y_2^2+3y_3^2)$ \\\cline{1-2}
\rule{0pt}{4ex} $A_3$ & $\dfrac{-4m_Wy_1^3y_2y_3}{2y_1^2-y_2^2}$ \\\cline{1-2}
\rule{0pt}{4ex} $A_4$ & $\dfrac{8m_Wy_1^3(y_2^2+y_3^2)}{2y_1^2-y_2^2}$ \\\cline{1-2}
\rule{0pt}{4ex} $A_5$ & $\dfrac{-4m_Wy_1^3}{2y_1^2-y_2^2}$ \\\cline{1-2}
\rule{0pt}{4ex} $A_6$ & $\dfrac{4m_Wy_1^3y_2}{y_3(2y_1^2-y_2^2)}$ \\\cline{1-2}
\end{tabular}
\label{feynrulesTabla4}
\end{table}

\end{document}